\documentclass[12pt,preprint]{aastex}

\begin{document}
\title{GAMMA-RAY LOUDNESS, SYNCHROTRON PEAK FREQUENCY, AND PARSEC-SCALE PROPERTIES OF BLAZARS DETECTED BY THE \textit{FERMI} LARGE AREA TELESCOPE}
\author{J. D. Linford, G. B. Taylor, and F. K. Schinzel}

\affil{University of New Mexico}
\affil{Department of Physics and Astronomy, MSC07 4220, Albuquerque, NM 87131-0001, USA}
\email{jlinford@unm.edu}



\begin{abstract}
The parsec-scale radio properties of 232 active galactic nuclei (AGNs), most of which are blazars, detected by the Large Area Telescope (LAT) on board the \textit{Fermi Gamma-ray Space Telescope} have been observed contemporaneously by the Very Long Baseline Array (VLBA) at 5 GHz.  Data from both the first 11 months (1FGL) and the first 2 years (2FGL) of the \textit{Fermi} mission were used to investigate these sources' $\gamma$-ray properties.  We use the ratio of the $\gamma$-ray to radio luminosity as a measure of $\gamma$-ray loudness.  We investigate the relationship of several radio properties to $\gamma$-ray loudness and to the synchrotron peak frequency.  There is a tentative correlation between $\gamma$-ray loudness and synchrotron peak frequency for BL Lac objects in both 1FGL and 2FGL, and for flat-spectrum radio quasars (FSRQs) in 2FGL.  We find that the apparent opening angle tentatively correlates with $\gamma$-ray loudness for FSRQs, but only when we use the 2FGL data.  We also find that the total VLBA flux density correlates with the synchrotron peak frequency for BL Lac objects and FSRQs.  The core brightness temperature also correlates with synchrotron peak frequency, but only for the BL Lac objects.  The low-synchrotron peaked (LSP) BL Lac object sample shows indications of contamination by FSRQs which happen to have undetectable emission lines.  There is evidence that the LSP BL Lac objects are more strongly beamed than the rest of the BL Lac object population.

\end{abstract}


\section{Introduction}
The Large Area Telescope (LAT; Atwood et al.\ 2009) on board the \textit{Fermi Gamma-ray Space Telescope} is a wide-field telescope covering the energy range from about 20 MeV to more than 300 GeV.  It has been scanning the entire $\gamma$-ray sky once every three hours since July of 2008, with breaks for flares and other targets of opportunity.  The majority of the sources (685 of 1451) in the \textit{Fermi} LAT First Source Catalog (1FGL; Abdo et al. 2010a) have been identified with known radio blazars.  These blazars typically are active galactic nuclei (AGNs) with strong, compact radio sources which exhibit flat radio spectra, rapid variability, compact cores with one-sided parsec-scale jets, and superluminal motion in the jets (Marscher 2006).  This trend continues in the newly released \textit{Fermi} LAT Second Source Catalog (2FGL; Nolan et al. 2011), compiled using the first two years of LAT data.

We previously presented findings on the relationships between the LAT-detected and non-LAT-detected populations of blazars (Linford et al. 2011; Linford et al. 2012).  Like other studies, we found a strong correlation between LAT flux and total VLBA radio flux density.  We also found that the LAT and non-LAT BL Lac objects appeared to be similar in many respects, while the LAT flat-spectrum radio quasars (FSRQs) were extreme sources when compared to their non-LAT counterparts.  Polarized emission at the base of the jets was also reported to be significantly more frequent in LAT blazars than in non-LAT blazars.

A major program to monitor the parsec-scale radio (15 GHz) properties of these $\gamma$-ray emitting blazars is the Monitoring Of Jets in AGN with VLBA Experiments (MOJAVE; Lister et al.\ 2009a; Homan et al.\ 2009).  The MOJAVE and \textit{Fermi} LAT collaborations recently published a paper detailing their investigations of parsec-scale properties of the $\gamma$-ray emitting blazars in their sample (Lister et al. 2011).  They studied the relationships between radio properties and the $\gamma$-ray loudness ($G_{r}$; the ratio of $\gamma$-ray-to-radio luminosity) and synchrotron peak frequency ($\nu^{S}_{peak}$; the frequency where the synchrotron emission is at maximum).  They reported significant differences in the $G_{r}$ distributions between the BL Lac objects and FSRQs.  For their BL Lac objects, they reported strong correlations for $G_{r}$-$\nu^{S}_{peak}$ and $G_{r}$-$\gamma$-ray photon spectral index.  They also reported a non-linear correlation between apparent jet opening angle and $G_{r}$ for their entire sample.  Their high-synchrotron peaked ($\nu^{S}_{peak} > 10^{15}$) BL Lac objects tended to have lower core brightness temperatures, linear core polarization, and variability than the rest of their BL Lac object population.

Here, we have analyzed our recent VLBA 5 GHz data, taken contemporaneously with LAT observations (Linford et al. 2012), to see if we could reproduce the findings of Lister et al. (2011).  Our sample was slightly larger than the one presented in Lister et al. (2011), with 232 sources in our sample compared to 173 in theirs.  We had a larger fraction of BL Lac objects in our sample, and we also had more non-blazar AGNs (radio galaxies, AGN of unknown type, and one starburst galaxy).  The objects in our sample spanned a larger range of radio flux densities, as the MOJAVE sample is targeting the brightest AGN and our sample is radio flux limited.  We only had single observations on our sources, as we are not a monitoring program.  This leads to some difficulties in comparing with the MOJAVE sample, especially in terms of apparent jet opening angle.  

Another area that has garnered renewed interest in the \textit{Fermi}-LAT era is the unification system for AGN (e.g., Urry \& Padovani 1995).  Several groups (e.g., Nieppola et al. 2008, Ghisellini \& Tavecchio 2008, Meyer et al. 2011, and Giommi et al. 2012b) have been investigating the ``blazar sequence'' (Fossati et al. 1998, Ghisellini et al. 1998), the relationship between blazar luminosity and synchrotron peak frequency.  There have been hints (Vermeulen et al. 1995, Ghisellini et al. 2009, Giommi et al. 2012a, Meyer et al. 2011) that the population of BL Lac objects with low (below 10$^{14}$ Hz) synchrotron peak frequencies might actually contain some misidentified FSRQs which happen to have strong emission from their jets overpowering their emission lines.  We investigated the possibility for this kind of ``contamination'' in our sample and present our findings here.

In Section 2 we define our sample.  In Section 3 we discuss how we determined the $\gamma$-ray loudness and synchrotron peak frequency for our sources. In Section 4 we present our results and discuss their implications.  In Section 5, we present evidence that our LSP BL Lac object sample may have some FSRQs hiding in it.  Throughout this paper we assume $\Lambda$CDM cosmology with $H_{0} = 71$ km s$^{-1}$ Mpc$^{-1}$, $\Omega_{m} = 0.27$, and $\Omega_{\Lambda} = 0.73$ (e.g., Hinshaw et al.\ 2009).

\section{Sample Definition}

The radio observations of our 232 sources were done with the Very Long Baseline Array (VLBA) at 5 GHz between 2009 November and 2010 July.  This is the sample described in more detail in Linford et al. (2012).  We included 90 sources that were follow-up observations of sources in the VLBA Imaging and Polarimetry Survey (VIPS; Helmboldt et al.\ 2007) as well as new 5 GHz observations of 7 MOJAVE sources.  The VIPS follow-up (or ``VIPS+'') sample can be thought of as radio-selected as the sources were originally selected and observed prior to the launch of \textit{Fermi}.  The remaining 135 sources (``VIPS++'') can be thought of as $\gamma$-ray-selected as they were specifically targeted due to their presence in 1FGL.  Where possible, we used the optical classifications from the LAT Second Catalog of AGN (2LAC; Ackermann et al. 2011).  If a source was not in 2LAC, we used the classification from the LAT First Catalog of AGN (1LAC; Abdo et al. 2010b).  In general, a source is classified as a BL Lac object if its strongest optical emission line has an equivalent width (EW) less than 5 \AA~ and the optical spectrum shows a CA II H/K break ratio of less than $0.4$.  An object is classified as a FSRQ if it has a flat radio spectrum and its optical spectrum is dominated by broad (EW $>5$ \AA) emission lines.  While we do not suspect widespread misclassification (such a problem would have a serious negative impact on our results and the results of every other study based on LAT data), we do present evidence that some objects classified as BL Lac objects may actually be FSRQs (see Section 5).  We should note that many sources classified in 1LAC as ``non-blazar AGN'' and ``AGN of unknown type'' were reclassified as BL Lac objects or FSRQs in 2LAC.  Therefore, the optical classifications for some of our objects are different than in Linford et al. (2012).  Our 1FGL sample contained 105 BL Lac objects, 114 FSRQs, and 13 other types of AGNs (radio galaxies, AGNs of unknown type, and 1 starburst galaxy).  Wherever possible, we used the 1LAC and 2LAC redshifts.  If a source did not have a redshift listed in 1LAC or 2LAC, we searched the NASA/IPAC Extragalactic Database (NED).

With the release of 2FGL and 2LAC, we also had newer data on several of our sources.  Unfortunately, not all of the sources in 1LAC were present in 2LAC, and some of our sources were among those dropped.  Our 2FGL sample contained 215 objects; 98 BL Lac objects, 108 FSRQs, and 9 AGN/Other.  Several 1FGL sources were dropped in 2FGL.  The 2FGL was based on 24 months of observing and some sources that were in a high-activity state (i.e., flaring) during the 1FGL period later decreased in brightness below the LAT threshold and their average flux failed to meet the criteria for inclusion in 2FGL.  For further discussion of the 1FGL-2FGL source comparisons, see Nolan et al. (2012).

While our sample is not statistically complete (i.e., it does not contain all northern hemisphere sources with a flux density greater than 30 mJy), it is representative of the LAT-detected blazar population.  It covers a wide range in flux density but is more biased towards relatively weak ($S < 1$ Jy) radio sources compared to the sample in Lister et al. (2011).  Because of this, we can expect to have fewer sources with low ($< 100$) $\gamma$-ray loudness (see Section 3.1) than in the sample presented in Lister at al. (2011).  Also, we are likely to have more BL Lac objects in our sample than other studies with a higher flux density threshold.  Our sample also covers a wide range of luminosities, synchrotron peak frequencies, and $\gamma$-ray loudness.  As Lister et al. (2011) discussed, representative samples are appropriate for statistical investigations and selection biases are not expected to have significant impacts on our results.  Obviously, the statistics of any study would be improved if a truly complete sample (i.e., simultaneous monitoring of all LAT-detected blazars observable by the VLBA) existed.  However, such a sample would require a prohibitive amount of VLBA observing time (in fact, it would probably have to be the only project observed for an entire year) to be practical.  Furthermore, a discussion of the biases introduced by our selection criteria would be, at best, speculation at this time as the complete parent sample of LAT sources has yet to be fully characterized.  The 2FGL catalog contains 577 sources with no known radio or optical association, which means we do not know what 31\% of the 2FGL sources are.  Also, of the 1121 entries in 2LAC, 270 are ``AGN of unknown type'', 11 are simply described as ``non-blazar AGN'', and 6 are listed as ``unidentified''.

The sample presented in Lister et al. (2011) is the combination of 2 statistically complete samples: a radio-selected sample and a $\gamma$-ray selected sample.  While the two samples on their own are complete, the combination is again representative.  Their sample contains more objects with high ($> 1$ Jy) flux density than ours, but they did not have as many sources with low ($< 500$ mJy) flux density.  Our sample contains more sources (232 versus 173), and in particular more BL Lac objects (105 versus 45), but our sample has fewer FSRQs (114 versus 123).

\section{Gamma-ray Loudness and Synchrotron Peak Frequency}

\subsection{Gamma-ray Loudness}
We adopted the definition of Lister et al. (2011) for ``$\gamma$-ray loudness'' as the ratio of the $\gamma$-ray luminosity to the radio luminosity.  Unlike Lister et al. (2011), we did not have multiple observations on our sources from which to determine a median radio luminosity.  Instead, we only used a single observation to calculate a luminosity.  Due to the variable nature of these blazars, this may not be the best representation of the sources' actual average luminosity.  For the $\gamma$-ray luminosity, we used the average fluxes reported in the 1FGL and the 2FGL.  The LAT measures $\gamma$-ray flux continuously from 20 MeV to over 300 GeV.  For publication (e.g., in 1FGL and 2FGL) the measurements are later grouped into energy bands in order to provide spectral information.  To obtain total $\gamma$-ray fluxes, we combined the fluxes from 3 bands where the LAT has the highest sensitivity: 100-300 MeV, 300 MeV - 1 GeV, and 1-100 GeV.  The fluxes were added and uncertainties were added in quadrature.  However, some sources had only upper limits to their fluxes in certain bands.  If a source's reported fluxes in one of the bands were upper limits, we use as the uncertainty half the reported flux in that band because the upper limits are given as 2-sigma results (Abdo et al. 2010a).  

To calculate the $\gamma$-ray luminosities, we used the same process as Lister et al. (2011).  We calculated the luminosities using data from both 1FGL/1LAC and 2FGL/2LAC, when available.  We started with converting energy fluxes using the equation
\begin{equation}
S_{0.1} = \frac{(\Gamma_{0}-1)C_{1}E_{1}F_{0.1}}{(\Gamma_{0}-2)}\left[1-\frac{E_{1}}{E_{2}}^{\Gamma_{0}-2}\right] \quad \textnormal{erg cm$^{-2}$ s$^{-1}$,}
\end{equation}
where $F_{0.1}$ is the LAT flux from 100 MeV to 100 GeV in photons cm$^{-2}$ s$^{-1}$, $E_{1}$ = 0.1 GeV, $E_{2}$ = 100 GeV, and $C_{1}$ = 1.602 $\times$ 10$^{-3}$ erg GeV$^{-1}$, and $\Gamma_{0}$ is the $\gamma$-ray photon spectral index which is set to 2.1 for this calculation. 
Next, the luminosity was calculated using the equation
\begin{equation}
L_{\gamma} = \frac{4 \pi D^{2}_{L} S_{0.1}}{(1+z)^{2-\Gamma}} \quad \textnormal{erg s$^{-1}$,}
\end{equation}
where $D_{L}$ is the luminosity distance\footnote{We calculated our $D_{L}$ values using a MATLAB\textregistered~ adaptation of Ned Wright's CosmoCalc program (Wright 2006).} in cm and $\Gamma$ is the $\gamma$-ray photon spectral index from 1LAC and 2LAC.

We calculated the radio luminosities using the equation
\begin{equation}
L_{R} = \frac{4 \pi D^{2}_{L} \nu S_{\nu}}{(1+z)} \quad \textnormal{erg s$^{-1}$,}
\end{equation}
where $\nu$ is 5 GHz, and $S_{\nu}$ is the total VLBA flux density we measured at 5 GHz.  We assumed a flat radio spectrum index ($\alpha$=0) for the purposes of the $k$-correction and luminosity calculations.

The $\gamma$-ray loudness is then simply
\begin{equation}
G_{r,1FGL} = \frac{L_{\gamma,1FGL}}{L_{R}}
\end{equation}
for 1FGL measurements and
\begin{equation}
G_{r,2FGL} = \frac{L_{\gamma,2FGL}}{L_{R}}
\end{equation}
for 2FGL measurements.

\begin{figure}
\plotone{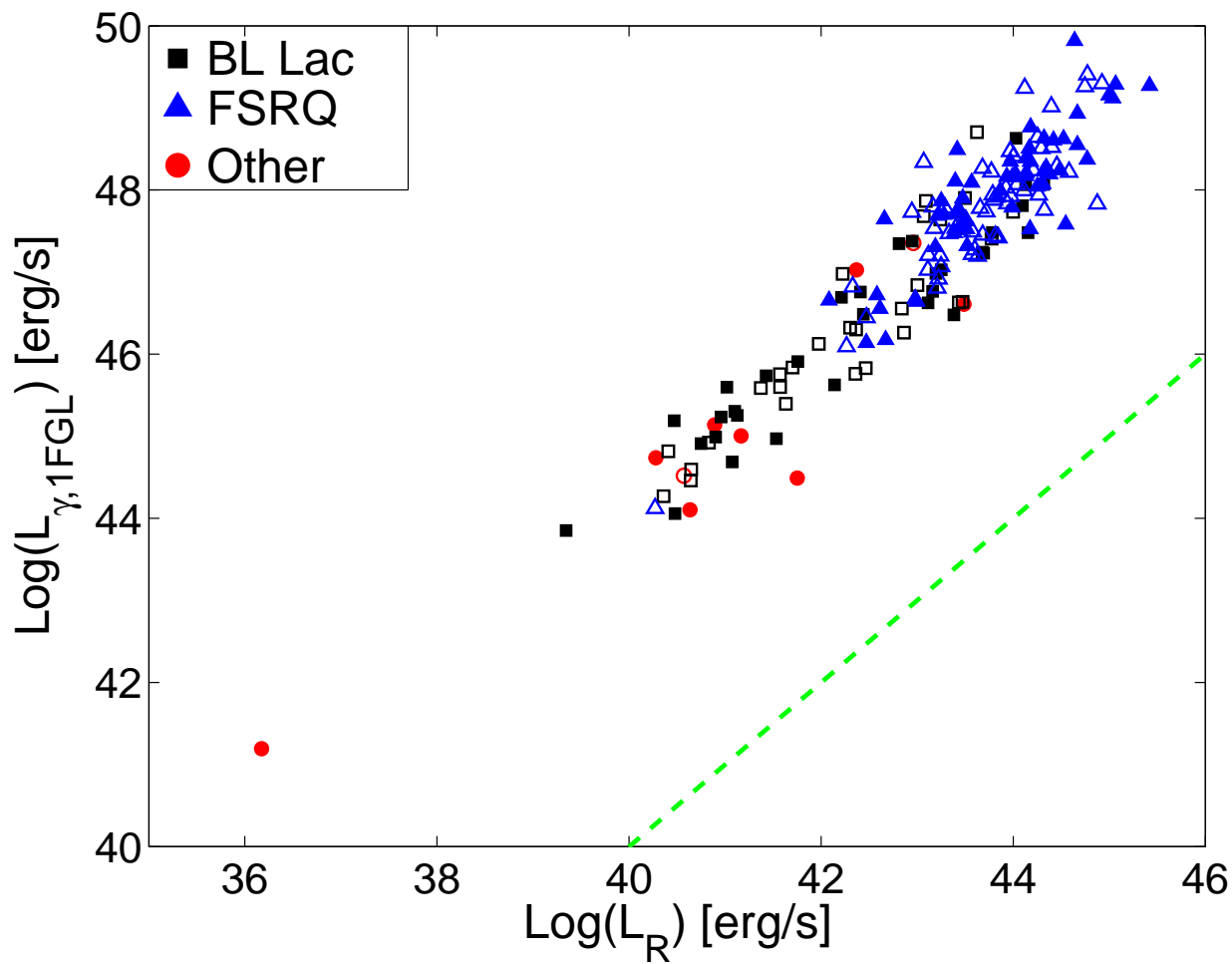}
\caption{1FGL $\gamma$-ray luminosity versus total VLBA luminosity at 5 GHz.  Black squares indicate BL Lac objects, blue triangles are FSRQs, and red circles are AGN/other.  Unfilled symbols are used for sources in VIPS+ and filled symbols are used for sources in VIPS++ (see Section 2). The dashed green line indicates a 1:1 luminosity ratio; any source above this line is considered $\gamma$-ray loud.  The ``other''-type source in the lower left is the starburst galaxy M82.  The 2FGL luminosity-luminosity plot is very similar.}
\label{ll_1fgl}
\end{figure}

All of our sources were significantly $\gamma$-ray loud.  For the 1FGL data, the minimum $G_{r,1FGL}$ was 545 and the maximum was 187000, with an average value of 19500.  For the 2FGL data, the minimum $G_{r,2FGL}$ was 404 and the maximum was 95100, with an average value of 13000. There are 2 possible explanations for this large change in maximum $G_{r}$ an the significant change in average $G_{r}$: \textit{i}) the 2FGL data were averaged over 2 years, so any sources in a high-activity/flaring state in the 11 months of 1FGL would naturally have lower $\gamma$-ray flux when averaged over a longer period of time, and \textit{ii}) the 1 - 100 GeV flux was calculated slightly differently for 2FGL than in 1FGL (Nolan et al. 2011).  We plot the 1FGL $\gamma$-ray versus radio luminosities in Figure~\ref{ll_1fgl}. The 2FGL luminosities were very similar to the 1FGL luminosities.

\begin{figure}
\plotone{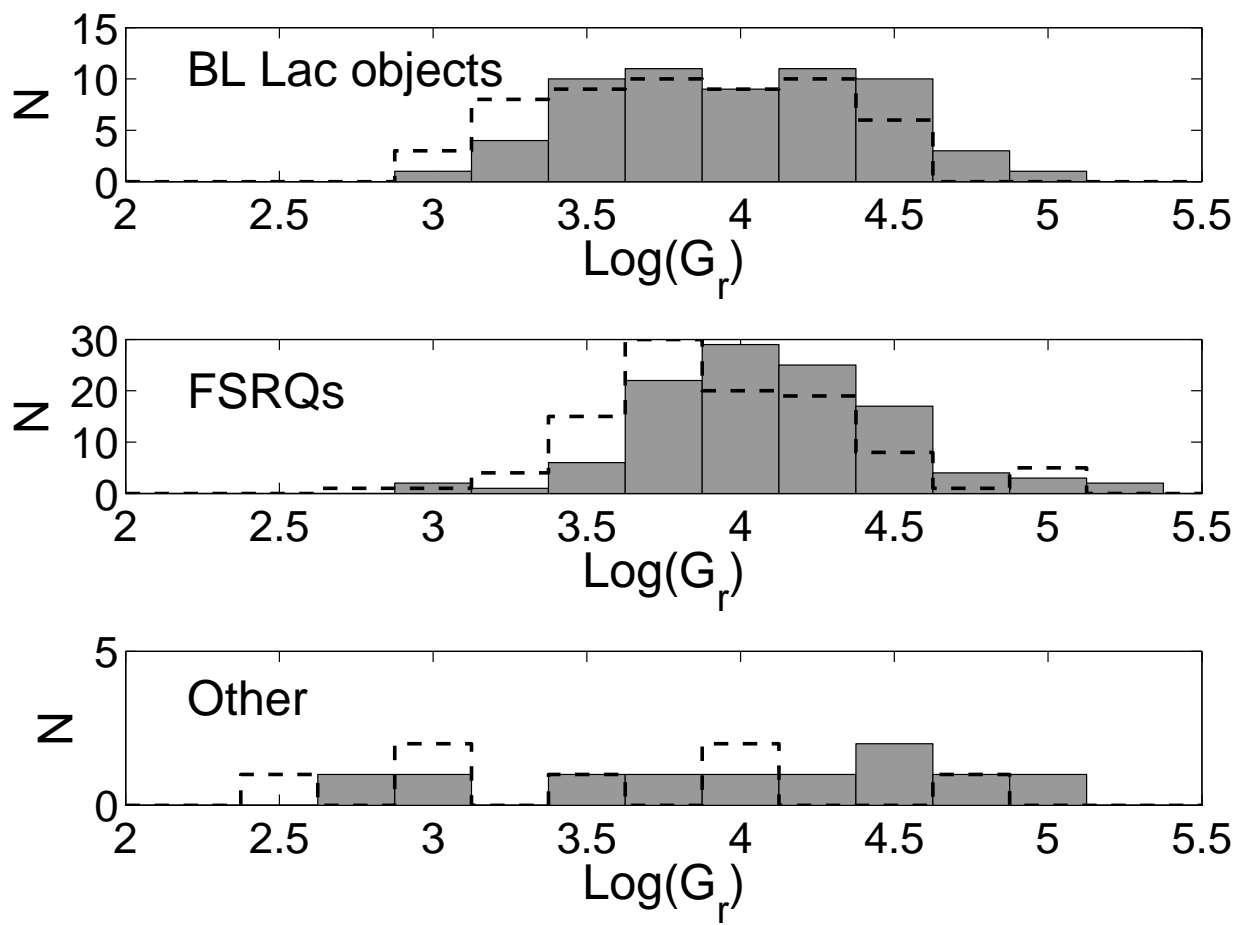}
\caption{Distributions of $\gamma$-ray to radio luminosity ratio for BL Lac objects (top), FSRQs (middle), and AGN/other (bottom).  The gray bars are for 1FGL data and the dashed lines are for 2FGL data.}
\label{G_hist_1}
\end{figure}

Lister et al. (2011), reported a significant difference between the distributions of $G_{r}$ for BL Lac objects and FSRQs.  We did not find such a difference in our data.  To examine the likelihood that the BL Lac objects and FSRQs were unrelated, we used the Kolmogorov-Smirnov (K-S) test\footnote{The K-S test is a useful statistic to determine the likelihood that two distributions are drawn from the same parent distribution.  It is important to remember that K-S test results are only meaningful in determining if two distributions are different.  That is, it should not be used to confirm that two distributions are similar, only that they are not drawn from the same parent population.} (e.g., Press et al. 1986).  The K-S test returned a 8\% probability that the 1FGL BL Lac objects and FSRQs were drawn from the same parent population.  For the 2FGL data, the K-S test result was 18\%.  See Figure~\ref{G_hist_1} for a plot of the $G_{r}$ distributions.

\subsection{Peak Synchrotron Frequency}

The frequency at which the synchrotron radiation is at a maximum ($\nu^{S}_{peak}$) is typically found by making many measurements of the spectrum of a source and fitting a polynomial to these measurements (e.g., Nieppola et al. 2006).  Ideally, one would like to have simultaneous multi-frequency measurements covering as much of the spectrum as possible.  Unfortunately, such large simultaneous data are rarely available.  When possible, we found estimates for $\nu^{S}_{peak}$ in the literature (see Table~\ref{datatable}).  It should also be noted that $\nu^{S}_{peak}$ is known to be variable (Rani et al. 2011).  Therefore, we made an effort to use the most recent published data.  We were able to find published $\nu^{S}_{peak}$ values for 102 of our sources, more than half of which came from Meyer et al. (2011).

If we could not find a published value of $\nu^{S}_{peak}$, we employed the technique used in 2LAC.  We used the average radio-optical ($\alpha_{RO}$) and optical-x-ray ($\alpha_{OX}$) spectral indices and the formula given in Abdo et al. (2010c).
\begin{equation}
\log(\nu^{S}_{peak}) = \left\{
\begin{array}{l l}
13.85 + 2.30X & \quad \textnormal{for $X<0$ and $Y<0.3$}\\
13.15 + 6.58Y & \quad \textnormal{otherwise}\\
\end{array} \right.
\end{equation}
Here, $X=0.565 - 1.433\alpha_{RO} + 0.155\alpha_{OX}$ and $Y = 1.0 - 0.661\alpha_{RO} - 0.339\alpha_{OX}$.  As per Giommi et al. (2012a), the $\nu^{S}_{peak}$ for FSRQs was reduced by 0.5 in $\log$ space.  Also, all $\nu^{S}_{peak}$ determined using the 2LAC method were expressed in the source rest frame.  For those BL Lac objects without a measured redshift, the median BL Lac object redshift of $0.27$ from the 2LAC was used, but only for the purposes of estimating $\nu^{S}_{peak}$ (i.e., we did not use this redshift to compute the luminosities).  We used the 2LAC estimation method to find $\nu^{S}_{peak}$ values for 76 of our sources.

If we could not find a published value and the $\alpha_{RO}$ and/or $\alpha_{OX}$ numbers were not in 2LAC, we used flux measurements from the NASA/IPAC Extragalactic Database (NED) and fit a quadratic to the data in $log(\nu F_{\nu}) - log(\nu)$ space.  Again, we used the most recent measurements available.  The $\nu^{S}_{peak}$'s found in this way are not as reliable as the 2LAC method, but they do tend to be reasonably close.  To check the accuracy of the NED-fit method, we made fits for all of our sources.  We threw out any fits that had less than 8 points for fitting and any that were obviously suspicious ($\nu^{S}_{peak,fit}<12$ and $\nu^{S}_{peak,fit}>19$), and then compared the remaining values to the published and 2LAC estimated values.  We calculated how close the $\nu^{S}_{peak,fit}$ was by using
\begin{equation}
\Delta_{\%} = \frac{\left|\nu^{S}_{peak} - \nu^{S}_{peak,fit}\right|}{\nu^{S}_{peak}}\times 100\%
\end{equation}
The mean for $\Delta_{\%}$ was $6.3\%$.  The median was $3.7\%$.  While this is not ideal, we feel it shows that the $\nu^{S}_{peak,fit}$'s are still acceptable estimates for those sources without a better alternative.  We were able to obtain $\nu^{S}_{peak,fit}$ estimates for 28 of our sources: 18 FSRQs, 8 BL Lac objects, and 2 AGN/other objects.  This brings our total number of sources with $\nu^{S}_{peak}$ values to 206.

As per the 1LAC convention, our sources were divided into 3 types based on their $\nu^{S}_{peak}$: High-synchrotron peaked (HSP, $\nu^{S}_{peak}>10^{15}$), Intermediate-synchrotron peaked (ISP, $10^{14}<\nu^{S}_{peak}<10^{15}$), and low-synchrotron peaked (LSP, $\nu^{S}_{peak}<10^{14}$).  For brevity, we will refer to the HSP BL Lac objects as ``HBLs'', ISP BL Lac objects as ``IBLs'', and LSP BL Lac objects as ``LBLs''.  We had 42 LBLs, 24 IBLs, and 29 LBLs in our sample.  We could not get good estimates of $\nu^{S}_{peak}$ for the remaining 11 BL Lac objects.  For the FSRQs, 99 were LSP type, and 3 were ISP type.  Notice that we do not have any HSP FSRQs in our sample.  We could not get reliable $\nu^{S}_{peak}$ estimates for 12 of our FSRQs.

\subsection{$G_{r}-\nu^{S}_{peak}$ Correlation}
\begin{figure}
\plotone{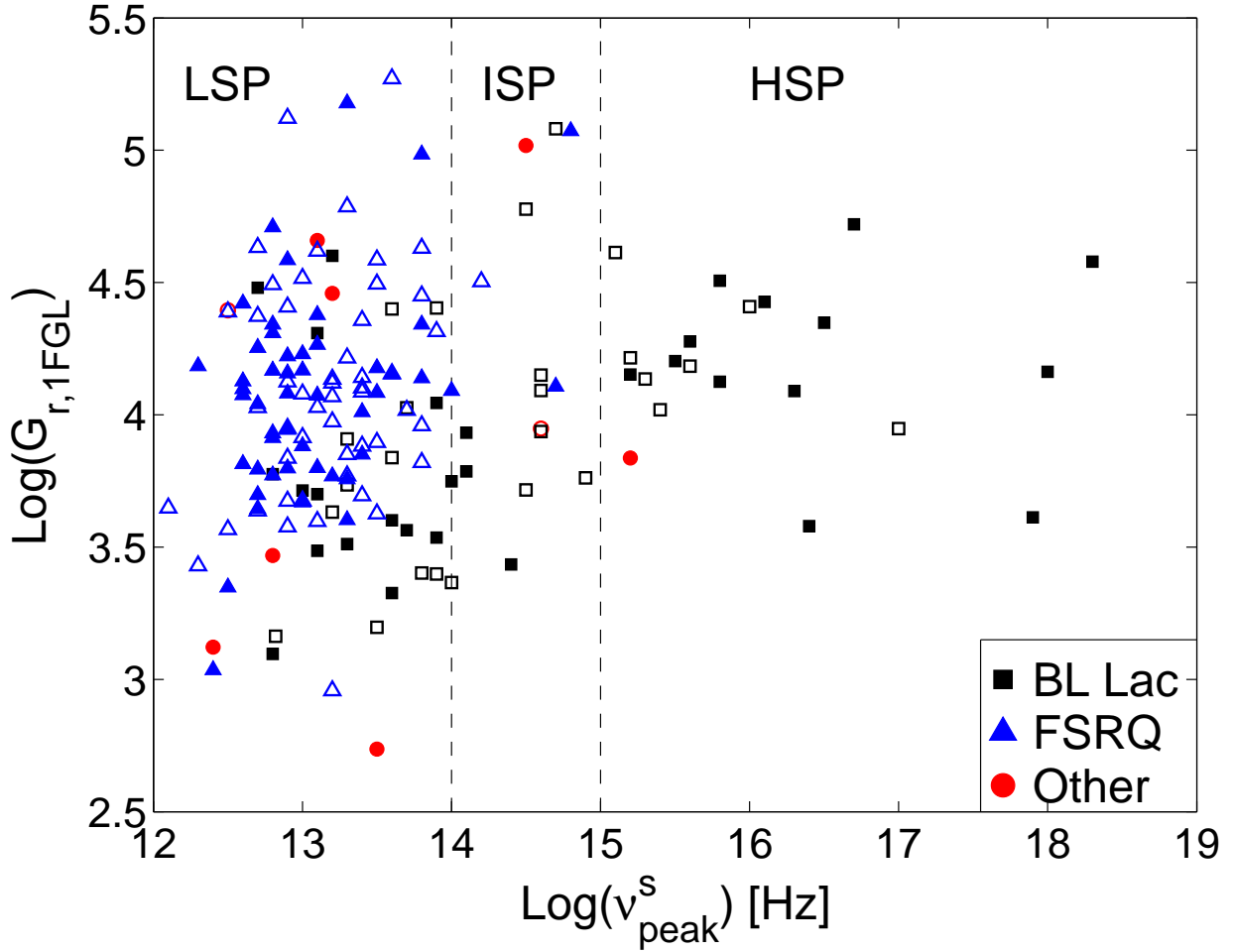}
\caption{1FGL $\gamma$-ray to radio luminosity ratio versus synchrotron peak frequency.  Unfilled symbols are used for sources in VIPS+ and filled symbols are used for sources in VIPS++ (see Section 2). The dashed lines indicate the divisions between low-, intermediate-, and high-synchrotron peaked objects.  The LSP ``other''-type object with low $G_{r}$ is the radio galaxy NGC 1275.  The ISP ``other''-type object with the relatively high $G_{r}$ is the starburst galaxy M82. The plot for $G_{r,2FGL}$ versus $\nu^{S}_{peak}$ is very similar.}
\label{G_nup_1}
\end{figure}

We applied the Spearman test\footnote{The nonparametric Spearman test returns a correlation coefficient ($\rho_{s}$), which has a range of $0<|\rho_{s}|<1$.  A high value of $|\rho_{s}|$ indicates a significant correlation.  The Spearman test also generates a significance ($p$).  The smaller the value of $p$, the less likely the chances of obtaining the same $\rho_{s}$ from random sampling.  It is important to keep in mind that while the Spearman test is a powerful test for statistical correlation, it does not test an actual physical correlation.} (e.g., Press et al. 1986) to look for correlations between $G_{r}$ and $\nu^{S}_{peak}$ in our data.  We found mostly weak or tentative correlations for our sample.  For the 1FGL data, the BL Lac objects had the only significant correlation with a $\rho_{S}$ of 0.46 and a $p$ of $4\times10^{-4}$.  The FSRQs in the 1FGL data had a very tentative correlation with a $\rho_{S}$ of 0.24 and a $p$ of $0.02$.  We plot our 1FGL data in Figure~\ref{G_nup_1}. For the 2FGL data, both the BL Lac objects and the FSRQs showed low-significance correlations.  The 2FGL BL Lac objects had a $\rho_{S}$ of 0.40 and a $p$ of 0.003 while the FSRQs had a $\rho_{S}$ of 0.28 and a $p$ of 0.006.

The 1FGL result tentatively confirms the correlation by Lister et al. (2011) who found a good correlation for the BL Lac objects, and no correlation for the FSRQs.  However, the  low-significance results from the 2FGL data cast some doubt on the correlation for the BL Lac objects and also call into question whether or not there is a correlation for the FSRQs.  Also, note that in Figure~\ref{G_nup_1} we have a handful of sources with high $\nu^{S}_{peak}$ and relatively low $G_{r}$.  With our large range in both, we do not suspect a selection effect here.  However, Lister et al. (2011) had significantly fewer HBLs than we had in our sample (17 versus 30), so their result of a strong, positive correlation may have been caused by a selection effect.  It does not seem to be caused by the fact that they were more biased towards high flux density objects.  We limited our BL Lac object sample to those brighter than 200 mJy and again to those brighter than 100 mJy, and in both cases our correlations became weaker than when we used our full sample.

A strong correlation between $G_{r}$ and $\nu^{S}_{peak}$ would indicate that the synchrotron self-Compton (SSC) model for $\gamma$-ray emission is more likely.  That is, the seed photons for the inverse Compton scattering to $\gamma$-ray frequencies are internal to the source (i.e., provided by the synchrotron emission) and therefore both the synchrotron and inverse Compton emission have roughly the same Doppler factors.  However, the fact that we see a lower significance correlation in the 2FGL data indicates that such a model may not be as accurate as previously thought.  The tentative correlation between $G_{r,2FGL}$ and $\nu^{S}_{peak}$ for the FSRQs could indicate that at least some fraction of their inverse Compton emission is SSC, but their seed photon population may be enriched by photons external to the synchrotron emitting region (i.e., external inverse Compton scattering) such as the dusty torus or broad line region (BLR).  Again, the low significance value for the correlation should make one cautious in drawing such a conclusion.
 
\section{Results and Discussion}

\subsection{Total VLBA Flux Density}
\begin{figure}
\plotone{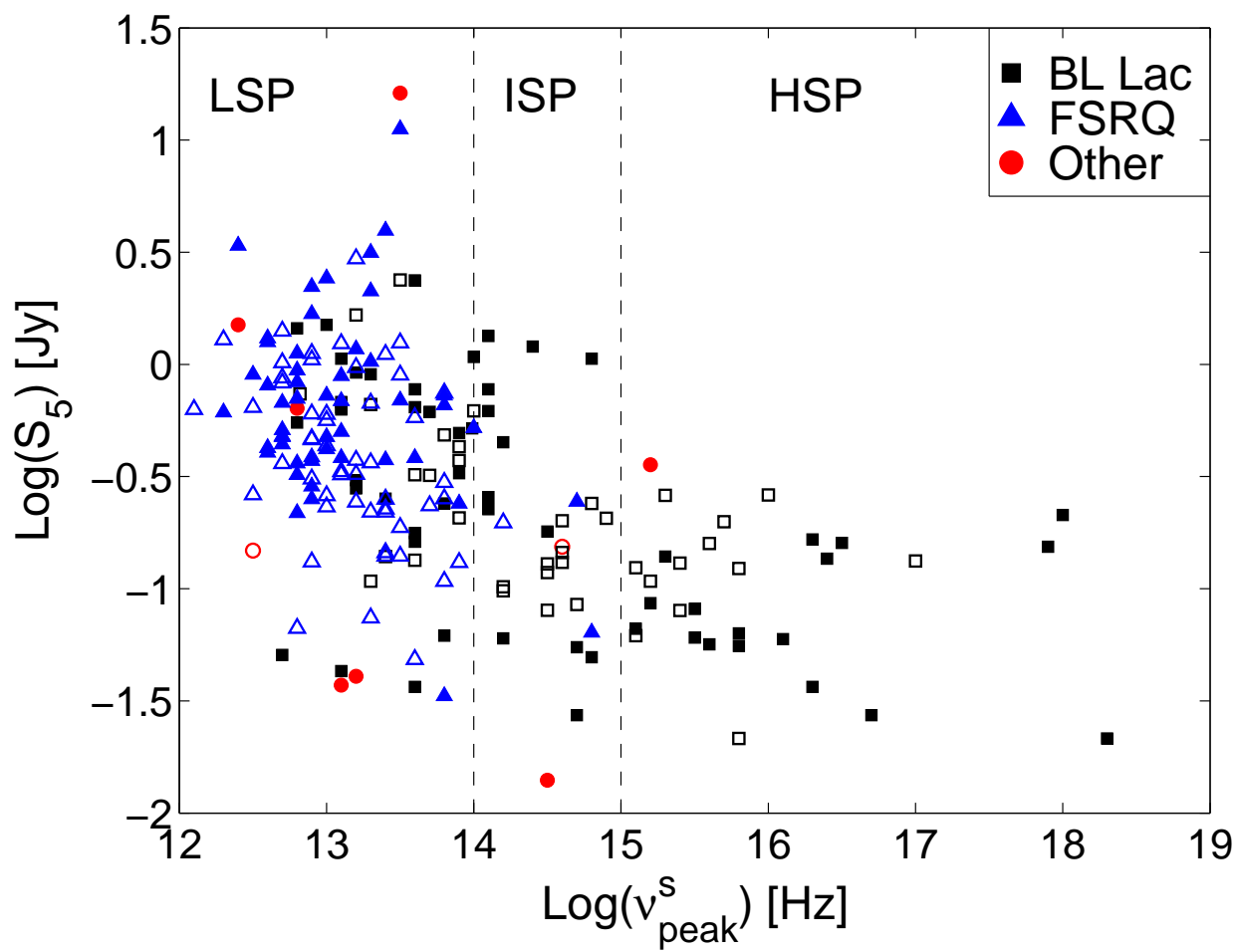}
\caption{Total VLBA flux density at 5 GHz versus synchrotron peak frequency.  Unfilled symbols are used for sources in VIPS+ and filled symbols are used for sources in VIPS++ (see Section 2).  The dashed lines indicate the divisions between low-, intermediate-, and high-synchrotron peaked objects. The LSP ``other''-type object with the high flux density is the radio galaxy NGC 1275.  The ISP ``other''-type object with the low flux density is the starburst galaxy M82.}
\label{S5_nup}
\end{figure}

We found very strong correlations between the total VLBA flux density at 5 GHz and the synchrotron peak frequency for our BL Lac objects and FSRQs.  The Spearman test result on the BL Lac objects was a $\rho_{S}$ of -0.56 and a $p$ of $5\times10^{-9}$.  The results for the FSRQs were a $\rho_{S}$ of -0.33 and a $p$ of $7\times10^{-4}$.  See Figure~\ref{S5_nup} for a plot of our flux density versus $\nu^{S}_{peak}$.  The $S_{5}-\nu^{S}_{peak}$ correlation is expected for the BL Lac objects and was also seen by Lister et al. (2011).  The higher the $\nu^{S}_{peak}$, the lower the radio flux density is going to be at 5 GHz which will contribute to a higher $G_{r}$.  The FSRQs were not expected to show such a strong correlation.  Lister et al. (2011) reported that their FSRQ sample showed no sign of correlation.  It is possible that, because the MOJAVE program focuses on the brightest AGN, their sample did not contain enough low flux density objects to show a correlation.  In fact, if we limit our FSRQ sample to those with $S_{5}>250$ mJy (excluding the 22 dimmest FSRQs with valid $\nu^{S}_{peak}$'s), we no longer see a significant correlation ($\rho_{S}=-0.04$ and $p=0.74$).

\subsection{Gamma-ray Photon Spectral Index}
\begin{figure}
\plotone{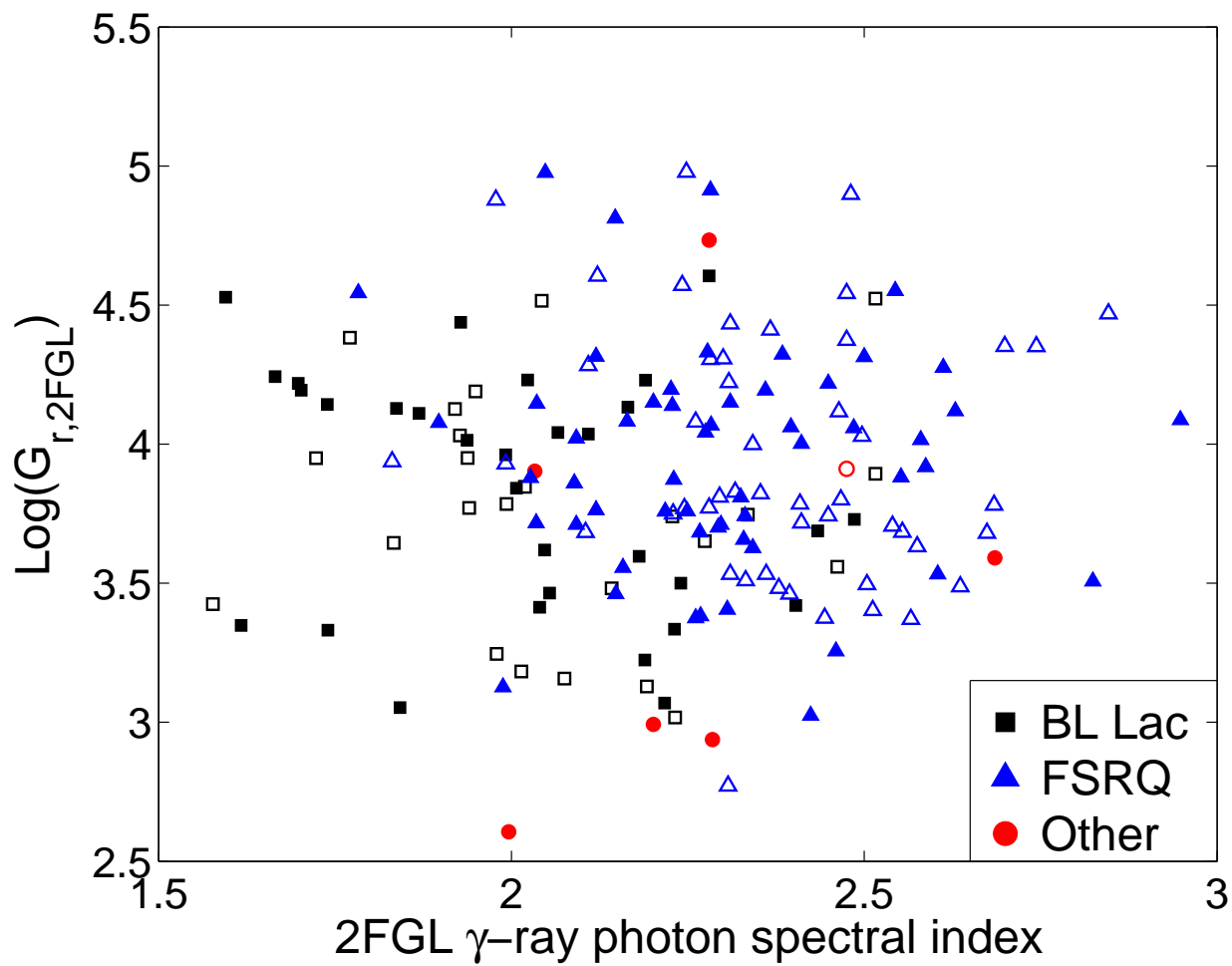}
\caption{$\gamma$-ray loudness versus $\gamma$-ray photon spectral index (all data from 2FGL).  Unfilled symbols are used for sources in VIPS+ and filled symbols are used for sources in VIPS++ (see Section 2).  The ``other''-type object with low $G_{r}$ and a photon spectral index of about 2 is the radio galaxy NGC 1275.}
\label{GA_Gr2}
\end{figure}

Lister et al. (2011) reported a significant linear correlation between the $\gamma$-ray loudness and $\gamma$-ray photon spectral index for their BL Lac objects.  We do not see a strong indication of this in our data.  We used both 1FGL and 2FGL data, and only the 2FGL data showed any hint of this correlation. However, with a $\rho_{S}$ of -0.25 and a $p$ of 0.06, it is still too low a significance to claim a correlation in our sample.  See Figure~\ref{GA_Gr2} for a plot of the 2FGL $\gamma$-ray loudness versus $\gamma$-ray photon spectral index.  We should note that while our sample covers a very large range in $G_{r}$, we do not have the low $G_{r}$ ($G_{r}<100$) objects that Lister et al. (2011) had in their sample.  It is possible that adding these low $G_{r}$ sources would lead to a strong correlation.

\begin{figure}
\plotone{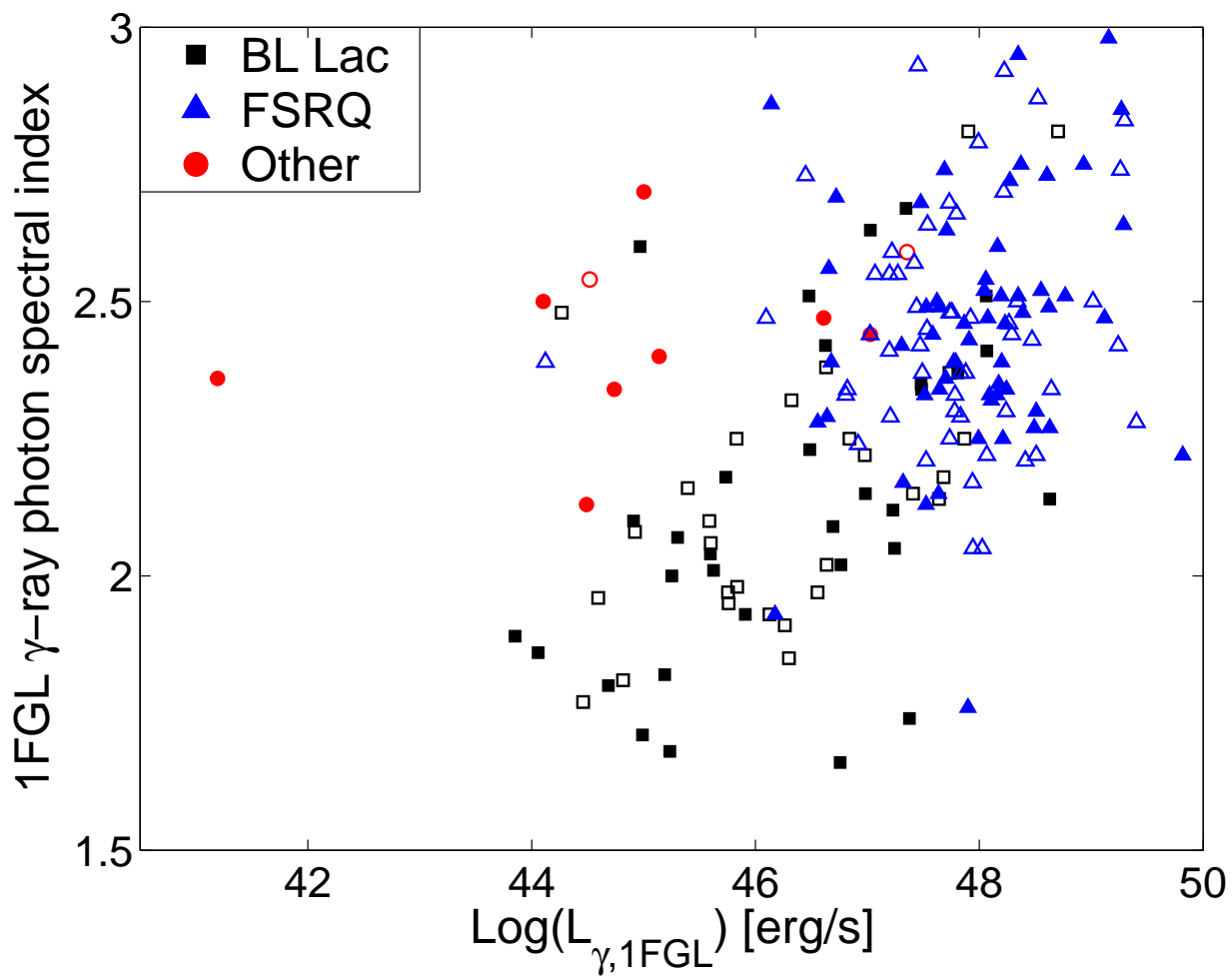}
\caption{$\gamma$-ray photon spectral index versus $\gamma$-ray luminosity (all data from 1FGL).  Unfilled symbols are used for sources in VIPS+ and filled symbols are used for sources in VIPS++ (see Section 2).  The ``other''-type object on the extreme left is the starburst galaxy M82.}
\label{LG_GA1}
\end{figure}

However, we do confirm the correlation between $\gamma$-ray luminosity and $\gamma$-ray photon spectral index reported by Ghisellini et al. (2009) and Chen \& Bai (2011).  For the 1FGL data, the Spearman test gives a $\rho_{S}$ of 0.43 and a $p$ of $2\times10^{-9}$ for the full sample.  The 2FGL data also showed a correlation with a $\rho_{S}$ of 0.35 and a $p$ of $4\times10^{-6}$.  Breaking the sources up by type, only the BL Lac objects show a correlation in either 1FGL ($\rho_{S} = 0.54$, $p = 8\times10^{-6}$) or 2FGL ($\rho_{S} = 0.50$, $p = 10^{-4}$).  See Figure~\ref{LG_GA1} for a plot of the 1FGL $\gamma$-ray photon spectral index versus $\gamma$-ray luminosity (the 2FGL plot looks similar and is not shown).  Ghisellini et al. (2009) and Chen \& Bai (2011) argue that this correlation indicates that the low-luminosity-low-$\nu^{S}_{peak}$ sources have lower black hole masses than the high-luminosity-high-$\nu^{S}_{peak}$ sources.  That is, the black hole mass, not beaming, may be responsible for the observed properties of low-luminosity LBLs.  However, our results from investigating the core brightness temperature and variability (below) do not support this.

\subsection{Core Brightness Temperature}
\begin{figure}
\plotone{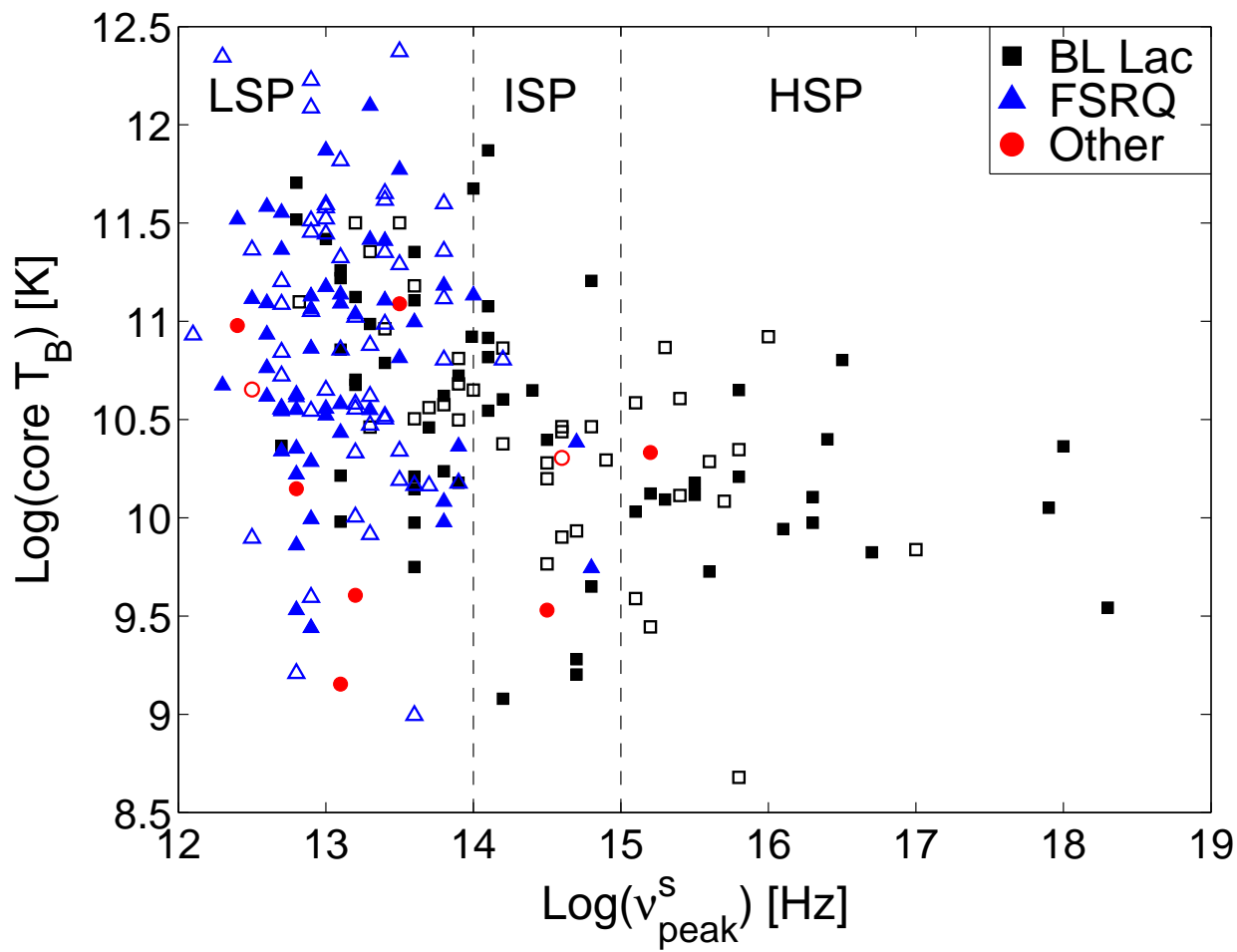}
\caption{Core brightness temperature versus synchrotron peak frequency.  Unfilled symbols are used for sources in VIPS+ and filled symbols are used for sources in VIPS++ (see Section 2).  The dashed lines indicate the divisions between low-, intermediate-, and high-synchrotron peaked objects.  The ``other''-type object with the highest core T$_{B}$ is the radio galaxy NGC 1275 (located in the LSP area).  The starburst galaxy M82 is the ISP ``other''-type object with the lower core T$_{B}$.}
\label{ct_nup}
\end{figure}

We found a correlation between the core brightness temperature and the peak synchrotron frequency for our sources.  When we applied the Spearman test to the entire sample, we found a $\rho_{S}$ value of -0.40 and a $p$ of $4\times10^{-9}$.  However, when we broke the sample up by optical type, only the BL Lac objects showed a significant correlation ($\rho_{S} = -0.55$, $p = 10^{-8}$).  See Figure~\ref{ct_nup} for a plot of the core brightness temperature versus $\nu^{S}_{peak}$.  High core brightness temperatures are generally associated with large Doppler factors (e.g., Tingay et al. 2001).  Therefore, it would seem that the LBLs are more strongly beamed than the HBLs.  Recall from Section 5.3 that the HBLs also tend to have larger $G_{r}$.  Combining their high $G_{r}$ and low core brightness temperatures indicates that the HBLs are probably more efficient at producing $\gamma$-rays.  The LBLs, on the other hand, may be seen as $\gamma$-ray loud thanks to higher Doppler factors.  Lister et al. (2011) did not report finding a correlation between core brightness temperature and $\nu^{S}_{peak}$, but they did note that their HBLs tended to have lower core brightness temperatures than the IBLs and LBLs.

\subsection{Apparent Jet Opening Angle}
We measured the mean apparent opening half angle for each source with core-jet morphology following the procedure described in Taylor et al. (2007).  This method relies on averaging the apparent opening angles found for each jet component in a source.  That is, the apparent opening angle is measured for several (more than 2) jet components.  The half-opening angle is calculated for each component to be 
\begin{equation}
\psi = {\rm arctan}[(|y'|+dr)/|x'|]
\end{equation}
where $x'$ and $y'$ are the positions of the component in a rotated coordinate system with the x'-axis aligned with the jet axis, and $dr$ is the deconvolved Gaussian size
perpendicular to the jet axis.  The final mean opening angle for the source is the average of all of the opening angles for the individual jet components.

\begin{figure}
\plotone{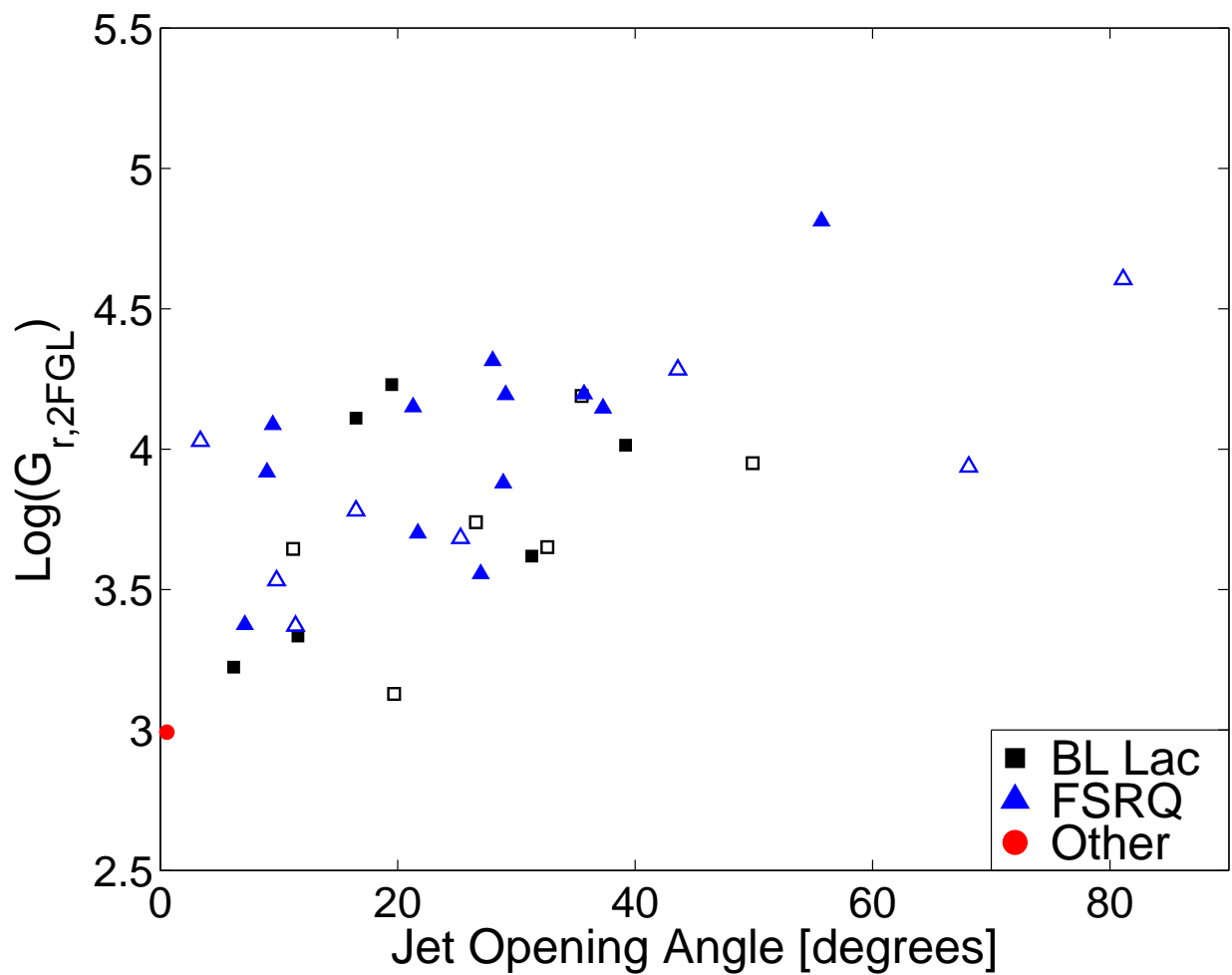}
\caption{2FGL $\gamma$-ray to radio luminosity ratio versus apparent jet opening angle.  Unfilled symbols are used for sources in VIPS+ and filled symbols are used for sources in VIPS++ (see Section 2).  The lone ``other''-type object is the radio galaxy NGC 6251.  The 1FGL data did not show a correlation.}
\label{G_oa_2}
\end{figure}

Lister et al. (2011) reported a non-linear correlation between the opening angle and the $\gamma$-ray loudness.  We found a tentative correlation, however only when we used the 2FGL data.  When we performed the Spearman test on all 33 sources with both a measured opening angle and a 2FGL $\gamma$-ray flux (including the radio galaxy NGC 6251), we found a $\rho_{S}$ of 0.57 and a $p$ of $5\times10^{-4}$.  When we break the sources up by type, we only see a tentative correlation for the FSRQs ($\rho_{S}$=0.62, $p$=0.004), however the sample sizes are very small.  We only had 36 objects with opening angle and $G_{r}$ measurements in 1FGL, and 33 in 2FGL.  The sample in Lister et al. (2011), on the other hand, contained well over 100 sources.  See Figure~\ref{G_oa_2} for a plot of the 2FGL $\gamma$-ray loudness versus opening angle.  We should also note that Lister et al. (2011) made their measurements of the apparent opening angle using mean sizes of jet components over several epochs, not the mean of multiple apparent opening angles from several components in a single observation as we did.  The fact that MOJAVE is a monitoring program makes it much better suited to investigating the apparent (and intrinsic) opening angles than our sample.

\subsection{Polarization}
To measure the polarization properties of our sources, we used the Gaussian mask method described by Helmboldt et al. (2007).  With this method, mask images were created using the Gaussian fit components in the Stokes $I$ image to define the core and jet components.  We then created mask images using the polarized intensity maps and mask out any pixels that do not have a signal-to-noise of at least 5 (compared to the noise image generated by the AIPS task COMB).  The two mask types were then combined (multiplied) and applied to both the Stokes $I$ and polarized intensity maps.  Additionally, in order to be considered ``polarized'', a source had to have a polarized flux of at least 0.3\% of the Stokes $I$ peak value (to avoid leakage contamination).  See Linford et al. (2012) for more discussion on the polarization poperties of our sources.

\begin{figure}
\plotone{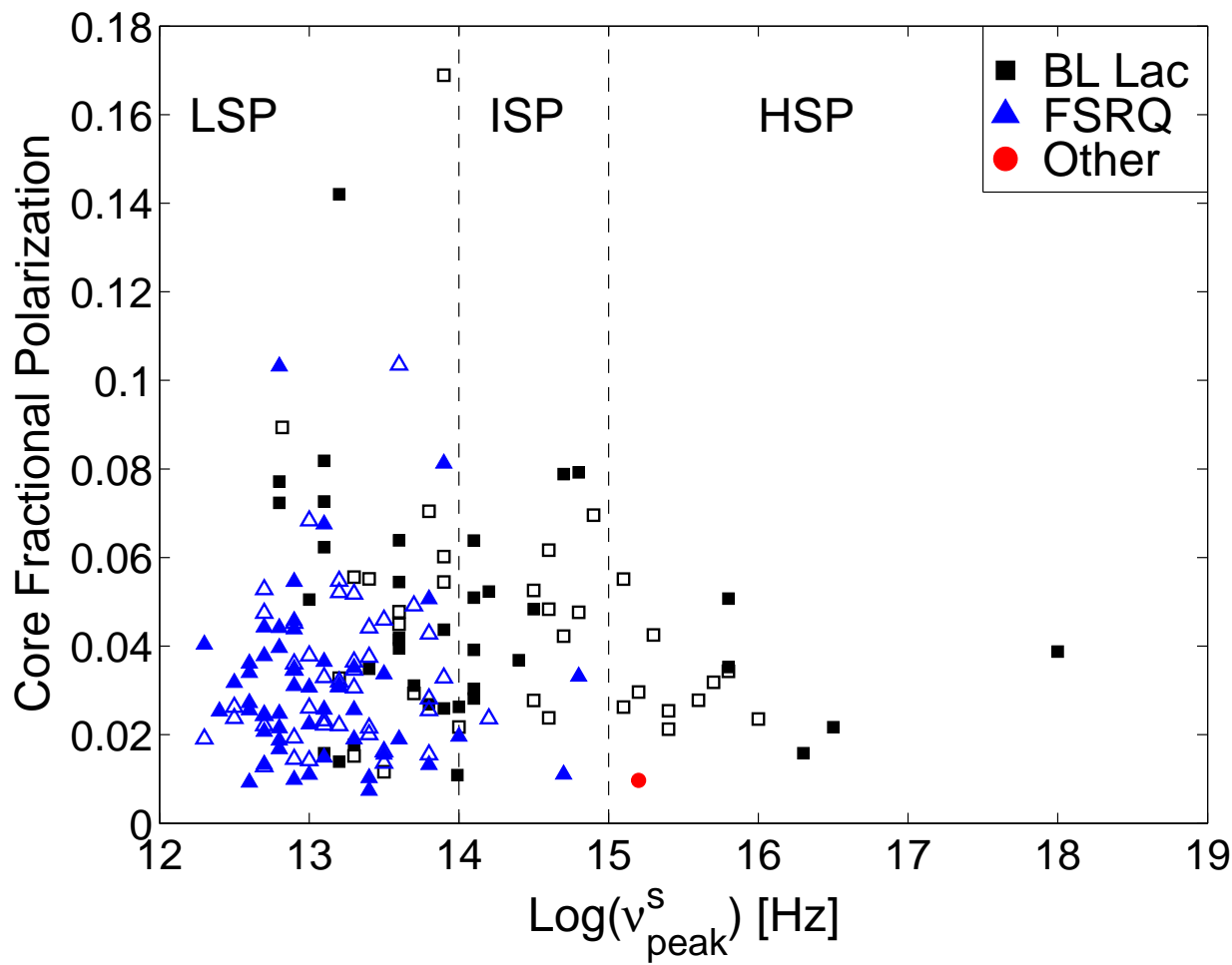}
\caption{5 GHz core fractional polarization versus synchrotron peak frequency.  Unfilled symbols are used for sources in VIPS+ and filled symbols are used for sources in VIPS++ (see Section 2).  The dashed lines indicate the divisions between low-, intermediate-, and high-synchrotron peaked objects. The lone ``other''-type object is the source F03250+3403, which has an uncertain classification: NED lists it as a Seyfert 1, 1LAC called it a non-blazar AGN, and 2LAC listed its type as ``unidentified''.}
\label{cfp_nup}
\end{figure}

Lister et al. (2011) reported that the HBLs in their sample tended to have lower core fractional polarization levels.  We also see an indication of this in our sample.  Our Spearman test returned only a very marginal correlation between core fractional polarization and $\nu^{S}_{peak}$ for the BL Lac objects, with a $\rho_{S}$ value of -0.25 and a $p$ of $0.04$.  However, we do note that the maximum HBL core fractional polarization is definitely less than the maximum for the IBLs or LBLs.  We did not find any correlation between core fractional polarization and $\nu^{S}_{peak}$ for the FSRQs (see Figure~\ref{cfp_nup} for a plot).

\subsection{Radio Variability}
\begin{figure}
\plotone{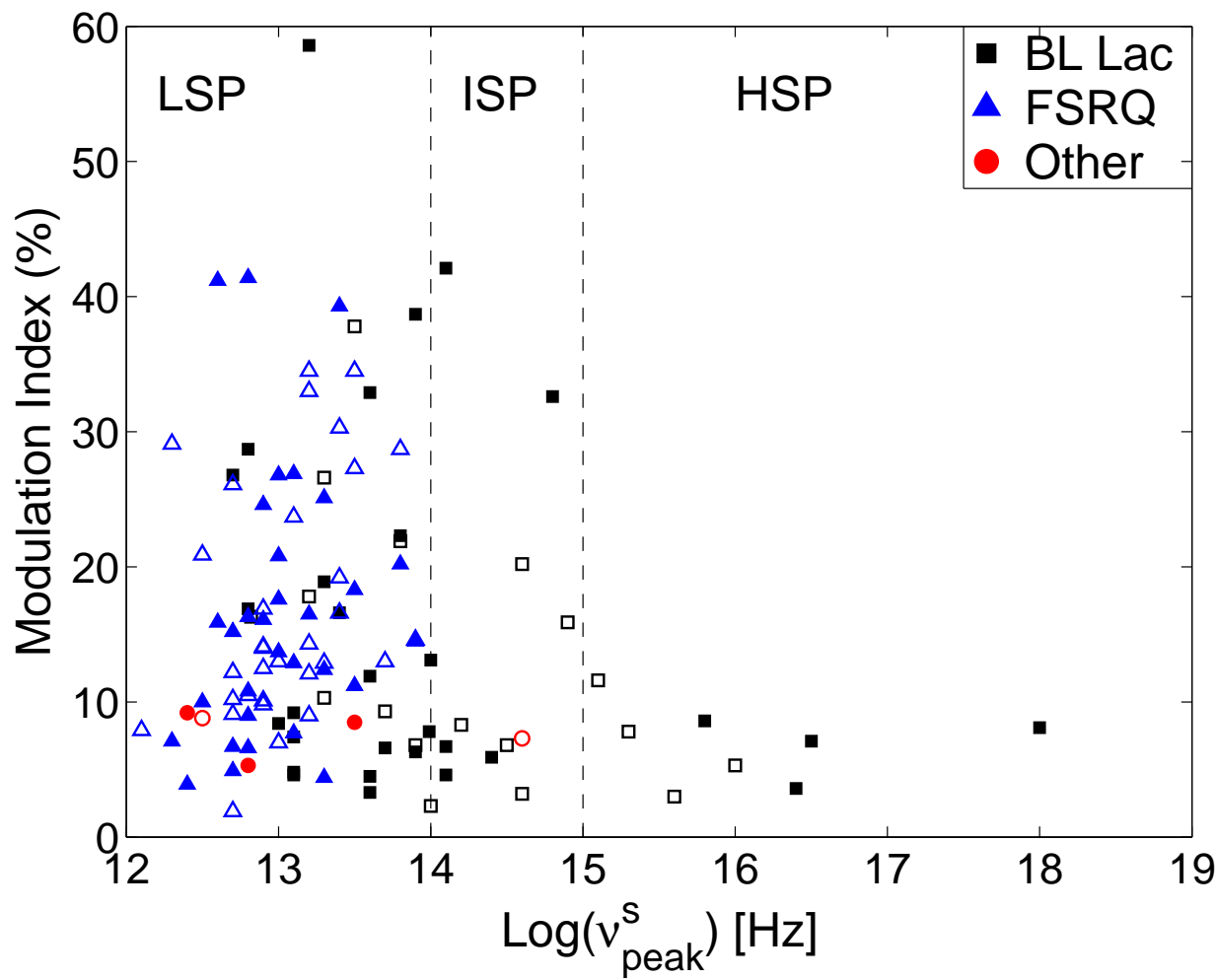}
\caption{Modulation index from Richards et al. (2011) versus synchrotron peak frequency.  Unfilled symbols are used for sources in VIPS+ and filled symbols are used for sources in VIPS++ (see Section 2).  The dashed lines indicate the divisions between low-, intermediate-, and high-synchrotron peaked objects.  The radio galaxies NGC 6251 and NGC 1275 are the LSP ``other''-type sources with $\nu^{S}_{peak}$ values near 12.8 and 13.5, respectively.}
\label{m_nup}
\end{figure}

It is well known that blazars tend to be highly variable sources.  It is believed that this variability is related to Doppler beaming (e.g., Hovatta et al. 2009) because small changes in bulk material velocity and/or orientation angle can lead to large changes in the observed flux.  It has also been shown that Doppler beaming can shorten the apparent timescales of flaring events (e.g., Lister 2001).  We used the modulation index from the Owens Valley Radio Observatory (OVRO) blazar monitoring program (Richards et al. 2011) as a measure of our sources variability.  The modulation index is the standard deviation of the distribution of source flux densities in time divided by the mean flux density.  Our sample and the OVRO sample had 120 sources in common.  Richards et al. (2011) reported that the FSRQs in their sample tended to have larger variability than the BL Lac objects.  We did not see strong evidence for this in our sample, but we should note that our sample is considerably smaller than Richards et al. (2011).  However, we did find that the HBLs tend to have relatively low variability (see Figure~\ref{m_nup}).  We also found a tentative correlation between the modulation index and $\nu^{S}_{peak}$ for the both the BL Lac objects and the FSRQs.  The Spearman test results for the BL Lac objects were a $\rho_{S}$ of -0.33 and a $p$ of 0.02.  This is another indication that the HBLs may not be as strongly beamed as the LBLs.  Interestingly, the FSRQs showed tentative positive correlation with a $\rho_{S}$ of 0.34 and a $p$ of 0.007.  However, it is hard to convince oneself that such a correlation exists for the FSRQs by examining Figure~\ref{m_nup}.

\subsection{Favoring 2LAC Estimates of Peak Synchrotron Frequency}
As we mentioned in Section 3.2, we opted to use the most recent published values of $\nu^{S}_{peak}$ when available.  However, it is likely that there were variations in how those values were determined in different studies.  An alternative to using recent published values is to use the 2LAC estimation method for as many sources as possible, and fill in the blanks with published and NED-fit values.  While the 2LAC method has been shown to be a reasonable estimate of $\nu^{S}_{peak}$, it is still an empirical estimate and does not always agree well with values calculated from fitting the SEDs.

We created a second data set favoring the 2LAC estimates of $\nu^{S}_{peak}$ and applied the Spearman test again to look for correlation.  For this secondary data set, we used the 2LAC estimated $\nu^{S}_{peak}$ values for 165 of our sources.  We also used 13 published values and 28 NED-fit values when the 2LAC estimation could not be calculated.  In general, the correlations reported above were still present, but with slightly reduced $\rho_{S}$ values (indicating weaker correlation) and increased $p$ values (indicating less significant correlation).  We did find a handful of significant differences when relying on the 2LAC values.  First, the FSRQs do not show any significant correlation for $G_{R}$ and $\nu^{S}_{peak}$ in either 1FGL or 2FGL.  Second, we no longer see any significant correlation between $\nu^{S}_{peak}$ and core fractional polarization for the BL Lac objects.  Finally, we no longer see any correlation between $\nu^{S}_{peak}$ and the modulation index (variability) for either BL Lac objects or FSRQs.  It should be noted that all the correlations that disappeared when relying on the 2LAC $\nu^{S}_{peak}$ estimates were marginal or tentative correlations using the published $\nu^{S}_{peak}$ values.

\section{Misidentified BL Lac Objects?}
It has recently been argued that some of the low-synchrotron BL Lac objects may not actually be BL Lac objects (e.g., Ghisellini et al. 2009, Giommi et al. 2012a).  In fact, it is possible that BL Lacertae itself is not actually a BL Lac object (Vermeulen et al. 1995).  The argument made is that some objects classified as LBLs are actually FSRQs with exceptionally strong jets and the broad line region (BLR) is simply not visible due to the jet emission overpowering the emission from the BLR (or, more familiarly, the jet is ``swamping'' the BLR).  The lack of obvious broad lines leads the astronomical community to mis-classify some sources as BL Lac objects.

\begin{figure}
\plotone{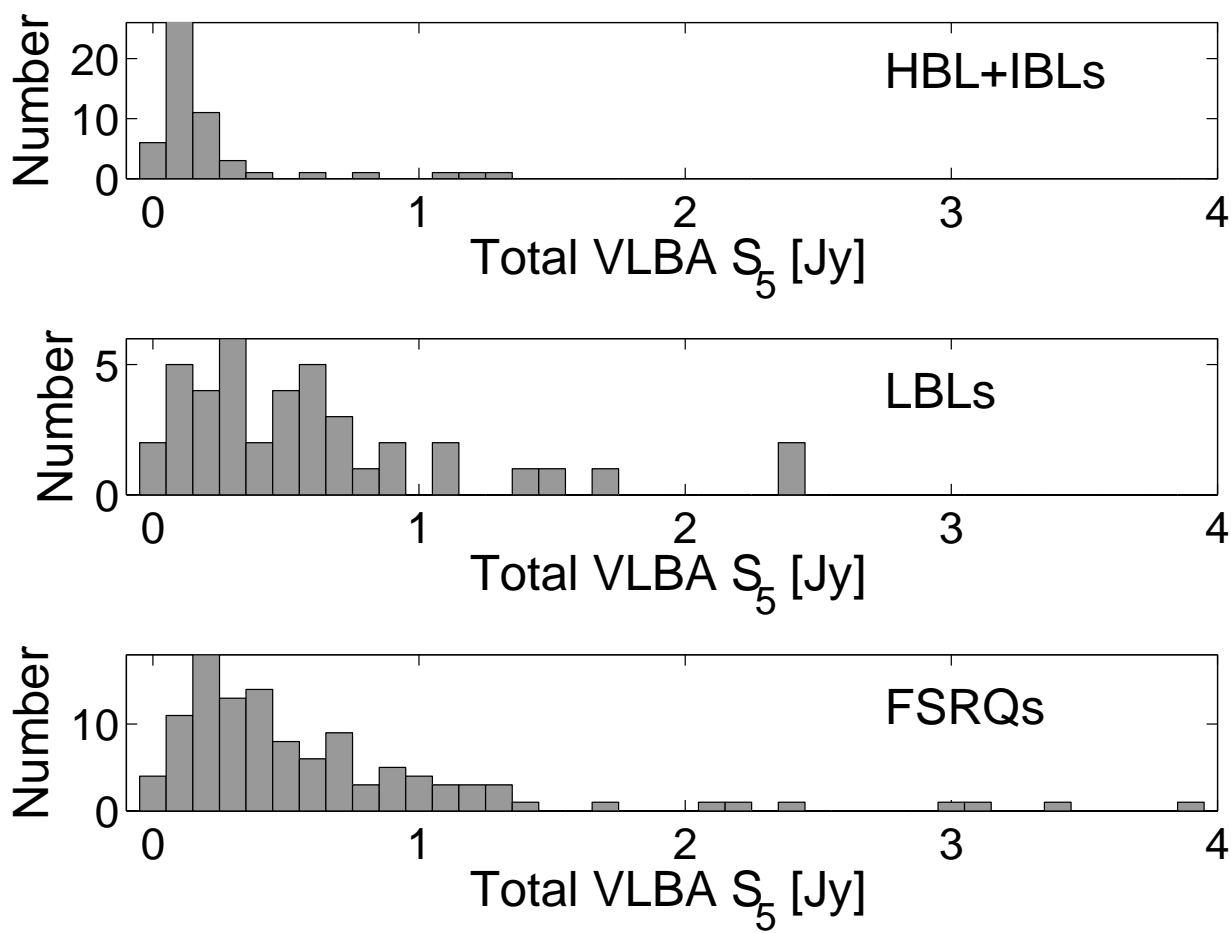}
\caption{Distributions of total VLBA flux density at 5GHz for the combined HSP and ISP BL Lac objects (top), LSP BL Lac objects (middle), and FSRQs (bottom).  We have omitted one FSRQ with a flux density of 11.2 Jy.}
\label{LBL_S5}
\end{figure}

\begin{figure}
\plotone{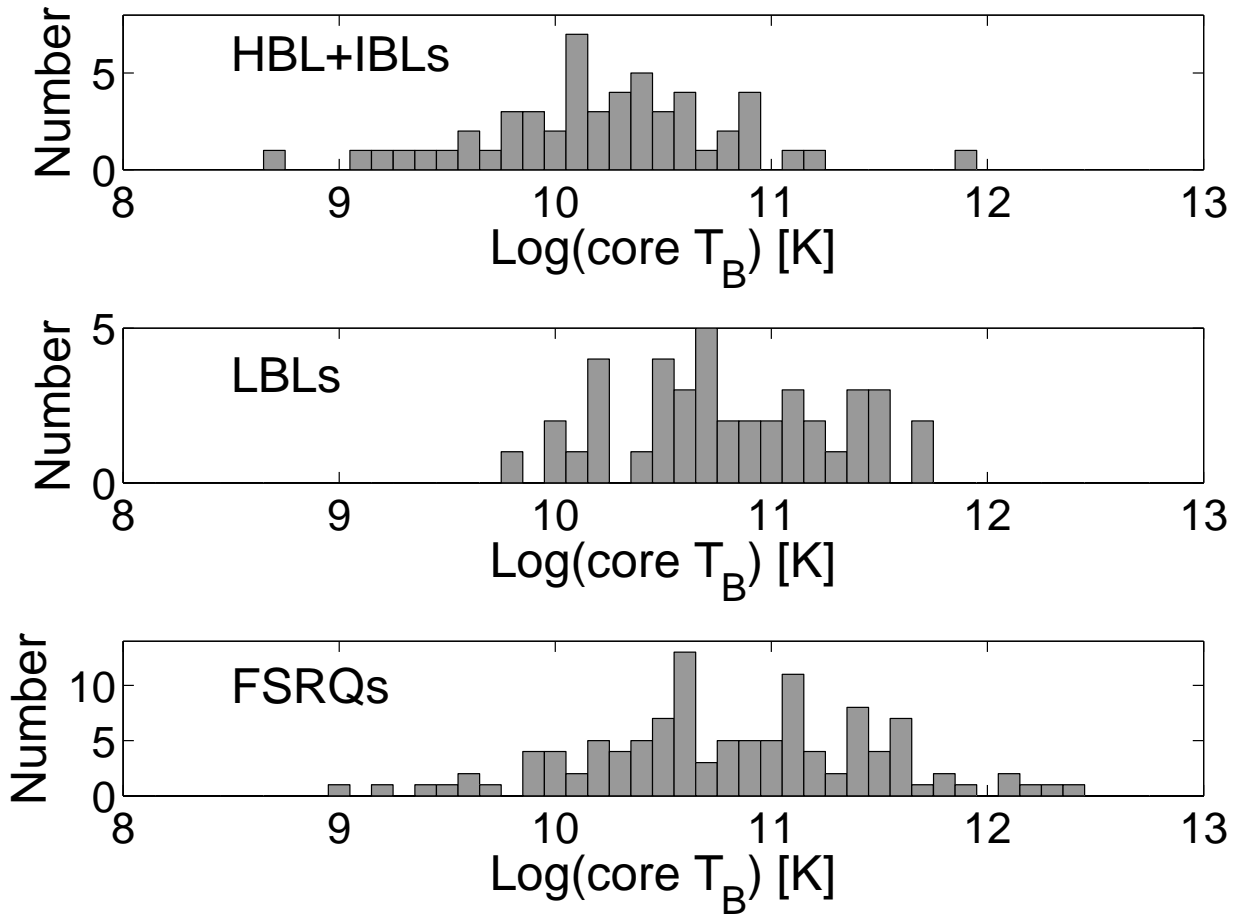}
\caption{Distributions of core brightness temperatures for the combined HSP and ISP BL Lac objects (top), LSP BL Lac objects (middle), and FSRQs (bottom).}
\label{LBL_Tb}
\end{figure}

To investigate this possibility, we compared the LBLs to the combined HBL+IBL population and to the FSRQs.  We looked at every parameter we had measured, and found that there are indeed several instances where the LBLs appear to be very different from the rest of the BL Lacs and are more like the FSRQs.  In particular, the distributions of the total VLBA radio flux density (see Figure~\ref{LBL_S5}) showed significant difference between the LBLs and the rest of the BL Lac objects.  We applied the K-S test to the distributions of total radio flux density and found that the probability of the LBLs and HBL+IBLs being drawn from the same parent sample was $6 \times 10^{-8}$.  The
core brightness temperatures for LBLs are also very unlike the core brightness temperatures of the rest of the BL Lac objects (see Figure~\ref{LBL_Tb}).  The K-S test result for the core brightness temperatures was a probability of $6 \times 10^{-5}$ that the LBLs and HBL+IBLs were drawn from the same parent sample.  

Interestingly, the K-S tests on the 1FGL and 2FGL $\gamma$-ray loudness distributions seem to indicate that the LBLs are not related to either the IBL+HBL or the FSRQ populations.  The apparent opening angle distributions were the only example where the LBLs showed strong likelihood of being unlike the FSRQs while not being very different from the IBL+HBL population.  However, this may simply be the result of low-number statistics.

Lister et al. (2011) found a difference between the core fractional polarization between the LBLs and HBLs.  We do not see any strong evidence for this in our sample. However, we should note that we do not use lower limits on our ``unpolarized'' sources, whereas the MOJAVE group did (Talvikki Hovatta, private communication).  Recall that we set ``unpolarized'' sources to have a core fractional polarization of 0\% and we only included sources with non-zero core fractional polarization in our ``polarized'' sample.  Therefore, we will naturally have different results than if we had included lower limits on those sources.

Lister et al. (2011) argued that their $G_{r} - \nu^{S}_{peak}$ correlation for BL Lac objects indicates that the LBLs and HBLs should belong to the same parent population.  However, we would argue that having some contamination of FSRQs masquerading as LBLs would not necessarily destroy any statistical correlation.  Furthermore, we did not find as strong of a $G_{r} - \nu^{S}_{peak}$ correlation for the BL Lac objects in our data.  To test our hypothesis that adding FSRQs to the LBL population would not destroy a statistical $G_{r} - \nu^{S}_{peak}$ correlation, we created contaminated samples by deliberately including known LSP FSRQs in our BL Lac object sample.  We then used the Spearman test to look for correlation in the contaminated samples.  Using 1FGL data, we found that after increasing the LBL sample size by 25\% with contaminating LSP FSRQs, the correlation results did not change significantly, and the correlation is still tentative after increasing the LBL sample size by 50\% and 100\%.  Using 2FGL data, the correlation becomes tentative after increasing the LBL sample size by 25\%, but it remains a tentative correlation even after increasing the LBL sample size by 50\% and 100\%.  See Table~\ref{phantomtable} for our full results.

While we cannot say for certain that we have some FSRQs masquerading as LBLs, it does seem likely.  Unfortunately, we cannot separate the misidentified BL Lac objects from the real ones just yet.  This will require monitoring these LBLs and watching for ones that have a drop in jet power leading to detection of a BLR (Vermeulen et al. 1995).

\section{Summary and Conclusions}
We have analyzed a sample of 232 LAT-detected AGN using both 1FGL and 2FGL data to compare our results with those of Lister et al. (2011).  All of the sources in our sample are significantly $\gamma$-ray loud.  We did not find a significant difference between the distributions of $G_{r}$ for BL Lac objects and FSRQs.  Using 1FGL data, we find a weak correlation between $G_{r}$ and $\nu^{S}_{peak}$ for the BL Lac objects and a tentative correlation for the FSRQs.  Using 2FGL, we found a very tentative $G_{r}$-$\nu^{S}_{peak}$ correlation for both FSRQs and BL Lac objects.  

Looking at the parsec-scale radio properties of our sources, we find a very strong correlation between total VLBA flux density and $\nu^{S}_{peak}$ for both BL Lac objects and FSRQs.  We could not confirm the correlation between $G_{r}$ and the $\gamma$-ray photon index reported by Lister et al. (2011), but we did confirm the correlation between $\gamma$-ray luminosity and $\gamma$-ray photon index reported by other groups.  We also found a very strong correlation between the core brightness temperatures and $\nu^{S}_{peak}$ for BL Lac objects.  Although we had a limited sample of apparent jet opening angle measurements, we were still able to tentatively confirm the correlation with $G_{r}$ reported by Lister et al. (2011).  We did not find any evidence of a correlation between core fractional polarization and $\nu^{S}_{peak}$.  We found a tentative negative correlation between radio variability (modulation index) and $\nu^{S}_{peak}$ for the BL Lac objects and a tentative positive correlation for the FSRQs.  The fact that the core brightness temperature shows a positive correlation with $\nu^{S}_{peak}$ and modulation index shows a negative correlation with $\nu^{S}_{peak}$ indicates that the LBLs are more strongly beamed than the IBLs and HBLs. 

The LBLs in our sample often appear to be different for the rest of the BL Lac objects.  In particular, we found significant differences in the distributions of core brightness temperatures and total VLBA flux density.  
It seems likely, therefore, that our LBL population contains some misidentified FSRQs which may have their BLR swamped by their jet emission.  While Lister et al. (2011) argued that a $G_{r}$-$\nu^{S}_{peak}$ correlation for the BL Lac objects indicated that the LBLs were related to the IBLs and HBLs, we found that deliberately contaminating our LBL sample with known FSRQs did not change our (albeit weak) correlation significantly.

Future studies of large samples of blazars, which should include both very high and low flux density objects, should be conducted to further investigate the relationships between $G_{r}$, $\nu^{S}_{peak}$, and the parsec-scale radio properties.  Long-term monitoring of LBLs may also present clear evidence that some of these objects are actually FSRQs.

\noindent
We thank the anonymous referee for their constructive criticism and helpful comments on this manuscript.
We thank Talvikki Hovatta for useful discussions regarding MOJAVE core fractional polarization measurements and Roger Romani for useful discussions about the 2LAC synchrotron peak frequency estimation method.  
The National Radio Astronomy Observatory is a facility of the National Science Foundation operated under cooperative agreement by Associated Universities, Inc.  
The VLBA is a facility of the National Science Foundation operated by the National Radio Astronomy Observatory under cooperative agreement by Associated Universities, Inc.
MATLAB is a registered trademark of The MathWorks, Inc. (Natick, Massachusetts, USA).
The NASA/IPAC Extragalactic Database (NED) is operated by the Jet Propulsion Laboratory, California Institute of Technology, under contract with the National Aeronautics and Space Administration.  
We thank NASA for support under FERMI grant GSFC \#21078/FERMI08-0051, and the NRAO for support under Student Observing Support Award GSSP10-011.  Additional support provided by the Naval Research Laboratory.

\textwidth = 7.0truein
\textheight = 10.0truein
\begin{deluxetable}{cccccccrr}
\tablecolumns{9}
\tabletypesize{\scriptsize}
\tablewidth{0pt}
\rotate
\tablecaption{Source Data}
\tablehead{
\colhead{Source Name} & \colhead{1FGL Name}	&	\colhead{2FGL Name}	&	\colhead{Alternate}	&	\colhead{Opt.}	&	\colhead{SED}	&	\colhead{SED}	&	\colhead{$G_{r,1FGL}$}	&	\colhead{$G_{r,2FGL}$} \\
\colhead{} & \colhead{} & \colhead{} &\colhead{AGN Name} & \colhead{Type} & \colhead{Type} & \colhead{Ref} & \colhead{} & \colhead{} }
\startdata
F00057+3815	&	 	1FGL J0005.7+3815	&	2FGL J0006.1+3821	&	 B2 0003+38A	&	 	bzq	&	LSP	&	2LAC	&	4675	&	3412	\\
F00193+2017	&	 	1FGL J0019.3+2017	&	\nodata	&	 PKS 0017+200	&	 	bzb	&	LSP	&	N-06	&	\nodata	&	\nodata	\\
F00230+4453	&	 	1FGL J0023.0+4453	&	2FGL J0023.2+4454	&	 B3 0020+446	&	 	bzq	&	\nodata	&	\nodata	&	41162	&	15615	\\
F00419+2318	&	 	1FGL J0041.9+2318	&	\nodata	&	 PKS 0039+230	&	 	bzq	&	LSP	&	fit	&	6226	&	\nodata	\\
F00580+3314	&	 	1FGL J0058.0+3314	&	2FGL J0057.9+3311	&	 CRATES J0058+3311	&	 	bzb	&	\nodata	&	\nodata	&	33541	&	13762	\\
F01022+4223	&	 	1FGL J0102.2+4223	&	2FGL J0102.3+4216	&	 CRATES J0102+4214	&	 	bzq	&	\nodata	&	\nodata	&	29901	&	18855	\\
F01090+1816	&	 	1FGL J0109.0+1816	&	2FGL J0109.0+1817	&	 CRATES J0109+1816	&	 	bzb	&	HSP	&	2LAC	&	13338	&	9149	\\
F01120+2247\tablenotemark{M}	&	 	1FGL J0112.0+2247	&	2FGL J0112.1+2245	&	 CGRaBS J0112+2244	&	 	bzb	&	LSP	&	M-11	&	11084	&	11018	\\
F01129+3207\tablenotemark{M}	&	 	1FGL J0112.9+3207	&	2FGL J0112.8+3208	&	 4C +31.03	&	 	bzq	&	LSP	&	2LAC	&	26411	&	14138	\\
F01138+4945	&	 	1FGL J0113.8+4945	&	2FGL J0113.7+4948	&	 CGRaBS J0113+4948	&	 	bzq	&	LSP	&	2LAC	&	4411	&	2373	\\
F01144+1327	&	 	1FGL J0114.4+1327\tablenotemark{*}	&	2FGL J0114.7+1326\tablenotemark{**}	&	 CRATES J0113+1324	&	 	bzb	&	\nodata	&	\nodata	&	34296	&	16980	\\
F01370+4751\tablenotemark{M}	&	 	1FGL J0137.0+4751	&	2FGL J0136.9+4751	&	 OC 457	&	 	bzq	&	LSP	&	M-11	&	5722	&	2894	\\
F01446+2703	&	 	1FGL J0144.6+2703	&	2FGL J0144.6+2704	&	 CRATES J0144+2705	&	 	bzb	&	LSP	&	2LAC	&	\nodata	&	\nodata	\\
F02035+7234	&	 	1FGL J0203.5+7234	&	2FGL J0203.6+7235	&	 CGRaBS J0203+7232	&	 	bzb	&	LSP	&	N-06	&	\nodata	&	\nodata	\\
F02045+1516	&	 	1FGL J0204.5+1516	&	2FGL J0205.0+1514	&	 4C +15.05	&	 	agn	&	LSP	&	M-11	&	1324	&	866	\\
F02053+3217\tablenotemark{M}	&	 	1FGL J0205.3+3217	&	2FGL J0205.4+3211	&	 B2 0202+31	&	 	bzq	&	LSP	&	2LAC	&	4006	&	1803	\\
F02112+1049	&	 	1FGL J0211.2+1049	&	2FGL J0211.2+1050	&	 CGRaBS J0211+1051	&	 	bzb	&	ISP	&	2LAC	&	\nodata	&	\nodata	\\
F02178+7353\tablenotemark{M}	&	 	1FGL J0217.8+7353	&	2FGL J0217.7+7353	&	 1ES 0212+735	&	 	bzq	&	LSP	&	M-11	&	7096	&	3217	\\
F02210+3555	&	 	1FGL J0221.0+3555	&	2FGL J0221.0+3555	&	 B2 0218+35	&	 	bzq	&	LSP	&	N-08	&	14719	&	11023	\\
F02308+4031	&	 	1FGL J0230.8+4031	&	2FGL J0230.8+4031	&	 B3 0227+403	&	 	bzq	&	LSP	&	fit	&	11845	&	13164	\\
F02379+2848	&	 	1FGL J0237.9+2848	&	2FGL J0237.8+2846	&	 4C +28.07	&	 	bzq	&	LSP	&	M-11	&	7632	&	3598	\\
F02386+1637	&	 	1FGL J0238.6+1637	&	2FGL J0238.7+1637	&	 PKS 0235+164	&	 	bzb	&	LSP	&	M-11	&	39868	&	17005	\\
F02435+7116	&	 	1FGL J0243.5+7116	&	2FGL J0242.9+7118	&	 CRATES J0243+7120	&	 	bzb	&	HSP	&	N-06	&	\nodata	&	\nodata	\\
F02454+2413	&	 	1FGL J0245.4+2413	&	2FGL J0245.1+2406	&	 B2 0242+23	&	 	bzq	&	LSP	&	2LAC	&	38503	&	35651	\\
F02580+2033	&	 	1FGL J0258.0+2033	&	2FGL J0257.9+2025\tablenotemark{c}	&	 CRATES J0258+2030	&	 	bzb	&	HSP	&	2LAC	&	\nodata	&	\nodata	\\
F03106+3812	&	 	1FGL J0310.6+3812	&	2FGL J0310.7+3813	&	 B3 0307+380	&	 	bzq	&	LSP	&	2LAC	&	10251	&	5742	\\
F03197+4130\tablenotemark{M}	&	 	1FGL J0319.7+4130	&	2FGL J0319.8+4130	&	 NGC 1275	&	 	agn	&	LSP	&	M-11	&	545	&	404	\\
F03250+3403	&	 	1FGL J0325.0+3403	&	2FGL J0324.8+3408	&	 B2 0321+33B	&	 	\nodata	&	HSP	&	2LAC	&	6861	&	3899	\\
F03259+2219	&	 	1FGL J0325.9+2219	&	2FGL J0326.1+2226	&	 CGRaBS J0325+2224	&	 	bzq	&	LSP	&	2LAC	&	18438	&	10045	\\
F03546+8009	&	 	1FGL J0354.6+8009	&	2FGL J0354.1+8010	&	 CRATES J0354+8009	&	 	agu	&	LSP	&	2LAC	&	\nodata	&	\nodata	\\
F04335+2905\tablenotemark{M}	&	 	1FGL J0433.5+2905	&	2FGL J0433.5+2905	&	 CGRaBS J0433+2905	&	 	bzb	&	LSP	&	2LAC	&	\nodata	&	\nodata	\\
F04335+3230	&	 	1FGL J0433.5+3230	&	2FGL J0433.7+3233	&	 CRATES J0433+3237	&	 	bzq	&	\nodata	&	\nodata	&	26569	&	35011	\\
F04406+2748	&	 	1FGL J0440.6+2748	&	2FGL J0440.9+2749	&	 B2 0437+27B	&	 	bzb	&	\nodata	&	\nodata	&	\nodata	&	\nodata	\\
F04486+112A	&	   1FGL J0448.6+1118\tablenotemark{*}	&	2FGL J0448.9+1121\tablenotemark{**}	&	 CRATES J0448+1127	&	  bzq	&	LSP	&	M-11	&	23877	&	20659	\\
F04486+112B	&	 	1FGL J0448.6+1118\tablenotemark{*}	&	2FGL J0448.9+1121\tablenotemark{**}	&	 PKS 0446+11	&	 	bzb	&	LSP	&	N-08	&	6522	&	5801	\\
F05092+1015	&	 	1FGL J0509.2+1015	&	2FGL J0509.2+1013	&	 PKS 0506+101	&	 	bzq	&	\nodata	&	\nodata	&	13083	&	6456	\\
F05100+180A	&			1FGL J0510.0+1800\tablenotemark{*}	&	2FGL J0509.9+1802	&	 CRATES J0509+1806	&	  agu	&	\nodata	&	\nodata	&	\nodata	&	\nodata	\\
F05100+180B	&	 	1FGL J0510.0+1800\tablenotemark{*}	&	2FGL J0509.9+1802	&	 PKS 0507+17	&	 	bzq	&	LSP	&	N-08	&	4989	&	5028	\\
F05310+1331\tablenotemark{M}	&	 	1FGL J0531.0+1331	&	2FGL J0530.8+1333	&	 PKS 0528+134	&	 	bzq	&	LSP	&	M-11	&	16680	&	5733	\\
F06072+4739	&	 	1FGL J0607.2+4739	&	2FGL J0607.4+4739	&	 CGRaBS J0607+4739	&	 	bzb	&	ISP	&	2LAC	&	\nodata	&	\nodata	\\
F06127+4120\tablenotemark{M}	&	 	1FGL J0612.7+4120	&	2FGL J0612.8+4122	&	 B3 0609+413	&	 	bzb	&	\nodata	&	\nodata	&	\nodata	&	\nodata	\\
F06169+5701	&	 	1FGL J0616.9+5701	&	2FGL J0616.9+5701	&	 CRATES J0617+5701	&	 	bzb	&	ISP	&	2LAC	&	\nodata	&	\nodata	\\
F06254+4440	&	 	1FGL J0625.4+4440	&	2FGL J0625.2+4441	&	 CGRaBS J0625+4440	&	 	bzb	&	LSP	&	N-06	&	\nodata	&	\nodata	\\
F06399+7325	&	 	1FGL J0639.9+7325	&	\nodata	&	 CGRaBS J0639+7324	&	 	bzq	&	LSP	&	M-11	&	15294	&	6293	\\
F06507+2503	&	 	1FGL J0650.7+2503	&	2FGL J0650.7+2505	&	 1ES 0647+250	&	 	bzb	&	HSP	&	N-06	&	37931	&	33759	\\
F06544+5042\tablenotemark{M}	&	 	1FGL J0654.4+5042	&	2FGL J0654.5+5043	&	 CGRaBS J0654+5042	&	 	agu	&	LSP	&	2LAC	&	\nodata	&	\nodata	\\
F06543+4514	&	 	1FGL J0654.3+4514	&	2FGL J0654.2+4514	&	 B3 0650+453	&	 	bzq	&	LSP	&	2LAC	&	51289	&	21437	\\
F07114+4731	&	 	1FGL J0711.4+4731	&	2FGL J0710.8+4733	&	 B3 0707+476	&	 	bzb	&	ISP	&	M-11	&	8555	&	5361	\\
F07127+5033	&	 	1FGL J0712.7+5033	&	2FGL J0712.9+5032	&	 CGRaBS J0712+5033	&	 	bzb	&	LSP	&	2LAC	&	\nodata	&	\nodata	\\
F07193+3306\tablenotemark{M}	&	 	1FGL J0719.3+3306	&	2FGL J0719.3+3306	&	 B2 0716+33	&	 	bzq	&	LSP	&	M-11	&	14239	&	11954	\\
F07219+7120\tablenotemark{M}	&	 	1FGL J0721.9+7120	&	2FGL J0721.9+7120	&	 CGRaBS J0721+7120	&	 	bzb	&	ISP	&	M-11	&	6111	&	6947	\\
F07253+1431	&	 	1FGL J0725.3+1431	&	2FGL J0725.3+1426	&	 4C +14.23	&	 	bzq	&	LSP	&	2LAC	&	6305	&	14005	\\
F07382+1741\tablenotemark{M}	&	 	1FGL J0738.2+1741	&	2FGL J0738.0+1742	&	 PKS 0735+178	&	 	bzb	&	LSP	&	M-11	&	3995	&	4161	\\
J07426+5444	&	 	1FGL J0742.2+5443	&	2FGL J0742.6+5442	&	 CRATES J0742+5444	&	 	bzq	&	LSP	&	2LAC	&	22776	&	16690	\\
J07464+2549	&	 	1FGL J0746.6+2548	&	2FGL J0746.6+2549	&	 B2 0743+25	&	 	bzq	&	LSP	&	2LAC	&	32807	&	29470	\\
F07506+1235\tablenotemark{M}	&	 	1FGL J0750.6+1235	&	2FGL J0750.6+1230	&	 PKS 0748+126	&	 	bzq	&	LSP	&	M-11	&	1087	&	1058	\\
J07530+5352	&	 	1FGL J0752.8+5353	&	2FGL J0753.0+5352	&	 4C +54.15	&	 	bzb	&	LSP	&	2LAC	&	2524	&	1522	\\
J08053+6144	&	 	1FGL J0806.2+6148	&	2FGL J0805.5+6145	&	 CGRaBS J0805+6144	&	 	bzq	&	LSP	&	2LAC	&	23584	&	22464	\\
J08096+3455	&	 	1FGL J0809.4+3455	&	\nodata	&	 B2 0806+35	&	 	bzb	&	\nodata	&	\nodata	&	6568	&	\nodata	\\
J08098+5218	&	 	1FGL J0809.5+5219	&	2FGL J0809.8+5218	&	 CRATES J0809+5218	&	 	bzb	&	HSP	&	M-11	&	16425	&	8916	\\
J08146+6431	&	 	1FGL J0815.0+6434	&	2FGL J0814.7+6429	&	 CGRaBS J0814+6431	&	 	bzb	&	ISP	&	2LAC	&	\nodata	&	\nodata	\\
J08163+5739	&	 	1FGL J0816.7+5739	&	2FGL J0816.5+5739	&	 BZB J0816+5739	&	 	bzb	&	HSP	&	M-11	&	\nodata	&	\nodata	\\
J08182+4222\tablenotemark{M}	&	 	1FGL J0818.2+4222	&	2FGL J0818.2+4223	&	 B3 0814+425	&	 	bzb	&	LSP	&	M-11	&	4287	&	3029	\\
J08247+5552	&	 	1FGL J0825.0+5555	&	2FGL J0824.9+5552	&	 OJ 535	&	 	bzq	&	LSP	&	2LAC	&	13192	&	6045	\\
J08308+2410\tablenotemark{M}	&	 	1FGL J0830.5+2407	&	2FGL J0830.5+2407	&	 OJ 248	&	 	bzq	&	LSP	&	M-11	&	7625	&	4792	\\
J08338+4224	&	 	1FGL J0834.4+4221	&	2FGL J0834.3+4221	&	 B3 0830+425	&	 	bzq	&	\nodata	&	\nodata	&	6741	&	3237	\\
F08422+7054\tablenotemark{M}	&	 	1FGL J0842.2+7054	&	2FGL J0841.6+7052	&	 4C +71.07	&	 	bzq	&	LSP	&	M-11	&	14357	&	12236	\\
F08499+4852	&	 	1FGL J0849.9+4852	&	2FGL J0849.8+4852	&	 CRATES J0850+4854	&	 	agu	&	ISP	&	2LAC	&	\nodata	&	\nodata	\\
J08548+2006\tablenotemark{M}	&	 	1FGL J0854.8+2006	&	2FGL J0854.8+2005	&	 OJ 287	&	 	bzb	&	LSP	&	M-11	&	1574	&	1041	\\
J08566+2057	&	 	1FGL J0856.6+2103\tablenotemark{*}	&	\nodata	&	 CRATES J0850+2057	&	 	bzq	&	\nodata	&	\nodata	&	56614	&	\nodata	\\
J08569+2111	&	 	1FGL J0856.6+2103\tablenotemark{*}	&	\nodata	&	 OJ 290	&	 	bzq	&	LSP	&	fit	&	16425	&	\nodata	\\
F09055+1356	&	 	1FGL J0905.5+1356	&	2FGL J0905.6+1357	&	 CRATES J0905+1358	&	 	agu	&	\nodata	&	\nodata	&	\nodata	&	\nodata	\\
J09106+3329	&	 	1FGL J0910.7+3332	&	2FGL J0910.6+3329	&	 Ton 1015	&	 	bzb	&	HSP	&	N-06	&	10457	&	5898	\\
J09121+4126	&	 	1FGL J0912.3+4127	&	2FGL J0912.1+4126	&	 B3 0908+416B	&	 	bzq	&	LSP	&	2LAC	&	25612	&	20286	\\
J09158+2933	&	 	1FGL J0915.7+2931	&	2FGL J0915.8+2932	&	 B2 0912+29	&	 	bzb	&	HSP	&	M-11	&	\nodata	&	\nodata	\\
J09209+4441\tablenotemark{M}	&	 	1FGL J0920.9+4441	&	2FGL J0920.9+4441	&	 B3 0917+449	&	 	bzq	&	LSP	&	A-11	&	42956	&	19211	\\
J09216+6215	&	 	1FGL J0919.6+6216	&	2FGL J0921.9+6216	&	 OK 630	&	 	bzq	&	LSP	&	M-11	&	4340	&	2528	\\
J09235+4125	&	 	1FGL J0923.2+4121	&	2FGL J0923.2+4125	&	 B3 0920+416	&	 	agn	&	LSP	&	fit	&	7086	&	3410	\\
J09238+2815	&	 	1FGL J0924.2+2812	&	2FGL J0924.0+2819	&	 B2 0920+28	&	 	bzq	&	LSP	&	fit	&	4228	&	3076	\\
J09292+5013	&	 	1FGL J0929.4+5000	&	2FGL J0929.5+5009	&	 CRATES J0929+5013	&	 	bzb	&	LSP	&	M-11	&	2503	&	1761	\\
J09341+3926	&	 	1FGL J0934.5+3929	&	2FGL J0934.7+3932	&	 CGRaBS J0934+3926	&	 	bzb	&	LSP	&	fit	&	8100	&	5579	\\
J09372+5008	&	 	1FGL J0937.7+5005	&	2FGL J0937.6+5009	&	 CGRaBS J0937+5008	&	 	bzq	&	LSP	&	2LAC	&	9429	&	3135	\\
J09418+2728	&	 	1FGL J0941.2+2722	&	2FGL J0941.4+2724\tablenotemark{**}	&	 CGRaBS J0941+2728	&	 	bzq	&	LSP	&	fit	&	10411	&	6734	\\
F09456+5754	&	 	1FGL J0945.6+5754	&	2FGL J0945.9+5751	&	 CRATES J0945+5757	&	 	bzb	&	LSP	&	fit	&	20374	&	13578	\\
F09466+1012	&	 	1FGL J0946.6+1012	&	2FGL J0946.5+1015	&	 CRATES J0946+1017	&	 	bzq	&	ISP	&	fit	&	12803	&	11436	\\
J09496+1752	&	 	1FGL J0949.8+1757\tablenotemark{*}	&	\nodata	&	 CRATES J0949+1752	&	 	bzq	&	\nodata	&	\nodata	&	8191	&	\nodata	\\
F09498+1757	&	 	1FGL J0949.8+1757\tablenotemark{*}	&	\nodata	&	 CRATES J0950+1804	&	 	agu	&	LSP	&	fit	&	45589	&	\nodata	\\
F09565+6938	&	 	1FGL J0956.5+6938	&	2FGL J0955.9+6936	&	 M 82	&	 	sbg	&	ISP	&	2LAC	&	104237	&	54147	\\
J09568+2515	&	 	1FGL J0956.9+2513	&	2FGL J0956.9+2516	&	 B2 0954+25A	&	 	bzq	&	LSP	&	M-11	&	3690	&	2897	\\
J09576+5522\tablenotemark{M}	&	 	1FGL J0957.7+5523	&	2FGL J0957.7+5522	&	 4C +55.17	&	 	bzq	&	LSP	&	M-11	&	14326	&	8662	\\
F10001+6539\tablenotemark{M}	&	 	1FGL J1000.1+6539	&	\nodata	&	 CGRaBS J0958+6533	&	 	bzb	&	LSP	&	A-11	&	1247	&	\nodata	\\
F10127+2440\tablenotemark{M}	&	 	1FGL J1012.7+2440	&	2FGL J1012.6+2440	&	 CRATES J1012+2439	&	 	bzq	&	ISP	&	0FGL	&	118387	&	94741	\\
J10150+4926\tablenotemark{M}	&	 	1FGL J1015.1+4927	&	2FGL J1015.1+4925	&	 1ES 1011+496	&	 	bzb	&	ISP	&	M-11	&	14129	&	8908	\\
J10330+4116	&	 	1FGL J1033.2+4116	&	2FGL J1033.2+4117	&	 B3 1030+415	&	 	bzq	&	LSP	&	M-11	&	2693	&	2370	\\
J10338+6051	&	 	1FGL J1033.8+6048	&	2FGL J1033.9+6050	&	 CGRaBS J1033+6051	&	 	bzq	&	LSP	&	2LAC	&	38586	&	37310	\\
F10377+5711\tablenotemark{M}	&	 	1FGL J1037.7+5711	&	2FGL J1037.6+5712	&	 CRATES J1037+5711	&	 	bzb	&	LSP	&	M-11	&	\nodata	&	\nodata	\\
J10431+2408	&	 	1FGL J1043.1+2404	&	2FGL J1043.1+2404	&	 B2 1040+24A	&	 	bzb	&	LSP	&	N-08	&	1455	&	1435	\\
F10487+8054	&	 	1FGL J1048.7+8054	&	\nodata	&	 CGRaBS J1044+8054	&	 	bzq	&	LSP	&	N-08	&	14738	&	\nodata	\\
F10485+7239	&	 	1FGL J1048.5+7239	&	2FGL J1049.7+7240	&	 CRATES J1047+7238	&	 	agu	&	\nodata	&	\nodata	&	\nodata	&	\nodata	\\
F10488+7145	&	 	1FGL J1048.8+7145	&	2FGL J1048.3+7144	&	 CGRaBS J1048+7143	&	 	bzq	&	LSP	&	2LAC	&	5867	&	4236	\\
J10586+5628\tablenotemark{M}	&	 	1FGL J1058.6+5628	&	2FGL J1058.6+5628	&	 CGRaBS J1058+5628	&	 	bzb	&	HSP	&	N-06	&	15262	&	10747	\\
J11044+3812\tablenotemark{M}	&	 	1FGL J1104.4+3812	&	2FGL J1104.4+3812	&	 Mkn 421	&	 	bzb	&	HSP	&	M-11	&	25649	&	24131	\\
J11061+2812	&	 	1FGL J1106.5+2809	&	2FGL J1106.1+2814	&	 CRATES J1106+2812	&	 	agu	&	LSP	&	2LAC	&	13858	&	5629	\\
J11126+3446	&	 	1FGL J1112.8+3444	&	2FGL J1112.4+3450	&	 CRATES J1112+3446	&	 	bzq	&	ISP	&	fit	&	31922	&	25768	\\
F11171+2013	&	 	1FGL J1117.1+2013	&	2FGL J1117.2+2013	&	 CRATES J1117+2014	&	 	bzb	&	HSP	&	2LAC	&	12285	&	15630	\\
J11240+2336	&	 	1FGL J1123.9+2339	&	2FGL J1124.2+2338	&	 OM 235	&	 	bzb	&	LSP	&	fit	&	\nodata	&	\nodata	\\
F11366+7009	&	 	1FGL J1136.6+7009	&	2FGL J1136.7+7009	&	 Mkn 180	&	 	bzb	&	HSP	&	M-11	&	3793	&	2143	\\
J11421+1547	&	 	1FGL J1141.8+1549	&	2FGL J1141.9+1550	&	 CRATES J1142+1547	&	 	agu	&	LSP	&	fit	&	\nodata	&	\nodata	\\
J11469+3958	&	 	1FGL J1146.8+4004	&	2FGL J1146.9+4000	&	 B2 1144+40	&	 	bzq	&	LSP	&	fit	&	8904	&	6644	\\
J11503+2417	&	 	1FGL J1150.2+2419	&	2FGL J1150.1+2419	&	 B2 1147+24	&	 	bzb	&	LSP	&	M-11	&	2323	&	1345	\\
J11514+5859	&	 	1FGL J1151.6+5857	&	2FGL J1151.5+5857	&	 CRATES J1151+5859	&	 	bzb	&	HSP	&	M-11	&	\nodata	&	\nodata	\\
J11540+6022	&	 	1FGL J1152.1+6027	&	2FGL J1154.4+6019	&	 CRATES J1154+6022	&	 	\nodata	&	LSP	&	M-11	&	\nodata	&	\nodata	\\
J11595+2914\tablenotemark{M}	&	 	1FGL J1159.4+2914	&	2FGL J1159.5+2914	&	 4C +29.45	&	 	bzq	&	LSP	&	M-11	&	7871	&	6459	\\
J12030+6031	&	 	1FGL J1202.9+6032	&	2FGL J1203.2+6030	&	 CRATES J1203+6031	&	 	agn	&	ISP	&	2LAC	&	12362	&	6094	\\
J12089+5441	&	 	1FGL J1209.3+5444	&	2FGL J1208.8+5441	&	 CRATES J1208+5441	&	 	bzq	&	LSP	&	fit	&	28188	&	23684	\\
J12093+4119	&	 	1FGL J1209.4+4119	&	2FGL J1209.6+4121	&	 B3 1206+416	&	 	bzb	&	ISP	&	N-06	&	8644	&	2661	\\
J12098+1810	&	 	1FGL J1209.7+1806	&	2FGL J1209.7+1807	&	 CRATES J1209+1810	&	 	bzq	&	LSP	&	fit	&	12218	&	13096	\\
J12178+3007\tablenotemark{M}	&	 	1FGL J1217.7+3007	&	\nodata	&	 B2 1215+30	&	 	bzb	&	HSP	&	M-11	&	13652	&	\nodata	\\
F12215+7106	&	 	1FGL J1221.5+7106	&	2FGL J1219.2+7107	&	 CRATES J1220+7105	&	 	bzq	&	\nodata	&	\nodata	&	3189	&	2413	\\
J12201+3431	&	 	1FGL J1220.2+3432	&	\nodata	&	 CGRaBS J1220+3431	&	 	bzb	&	ISP	&	N-06	&	5198	&	\nodata	\\
J12215+2813\tablenotemark{M}	&	 	1FGL J1221.5+2814	&	2FGL J1221.4+2814	&	 W Com	&	 	bzb	&	LSP	&	M-11	&	10627	&	7027	\\
F12248+8044	&	 	1FGL J1224.8+8044	&	2FGL J1223.9+8043	&	 CRATES J1223+8040	&	 	bzb	&	ISP	&	N-06	&	\nodata	&	\nodata	\\
J12248+4335	&	 	1FGL J1225.8+4336\tablenotemark{*}	&	2FGL J1225.0+4335\tablenotemark{**}	&	 B3 1222+438	&	 	agu	&	LSP	&	fit	&	25354	&	7824	\\
J12249+2122\tablenotemark{M}	&	 	1FGL J1224.7+2121	&	2FGL J1224.9+2122	&	 4C +21.35	&	 	bzq	&	LSP	&	M-11	&	6612	&	40266	\\
J12269+4340	&	 	1FGL J1225.8+4336\tablenotemark{*}	&	2FGL J1225.0+4335\tablenotemark{**}	&	 B3 1224+439	&	 	bzq	&	ISP	&	fit	&	120486	&	33357	\\
J12302+2518	&	 	1FGL J1230.4+2520	&	2FGL J1230.2+2517	&	 ON 246	&	 	bzb	&	ISP	&	N-06	&	5782	&	4484	\\
F12316+2850	&	 	1FGL J1231.6+2850	&	2FGL J1231.7+2848	&	 B2 1229+29	&	 	bzb	&	HSP	&	2LAC	&	14203	&	12907	\\
F12431+3627	&	 	1FGL J1243.1+3627	&	2FGL J1243.1+3627	&	 B2 1240+36	&	 	bzb	&	HSP	&	M-11	&	26724	&	16535	\\
J12483+5820\tablenotemark{M}	&	 	1FGL J1248.2+5820	&	2FGL J1248.2+5820	&	 CGRaBS J1248+5820	&	 	bzb	&	HSP	&	N-06	&	41029	&	15463	\\
J12531+5301	&	 	1FGL J1253.0+5301	&	2FGL J1253.1+5302	&	 CRATES J1253+5301	&	 	bzb	&	ISP	&	N-06	&	\nodata	&	\nodata	\\
J12579+3229	&	 	1FGL J1258.3+3227	&	2FGL J1258.2+3231	&	 B2 1255+32	&	 	bzq	&	LSP	&	2LAC	&	4721	&	4833	\\
J13030+2433\tablenotemark{M}	&	 	1FGL J1303.0+2433	&	2FGL J1303.1+2435	&	 CRATES J1303+2433	&	 	bzb	&	LSP	&	fit	&	25134	&	13368	\\
F13060+7852	&	 	1FGL J1306.0+7852	&	2FGL J1305.7+7854	&	 CRATES J1305+7854	&	 	agu	&	\nodata	&	\nodata	&	\nodata	&	\nodata	\\
J13083+3546	&	 	1FGL J1308.5+3550	&	2FGL J1308.5+3547	&	 CGRaBS J1308+3546	&	 	bzq	&	LSP	&	2LAC	&	13335	&	5903	\\
F13092+1156	&	 	1FGL J1309.2+1156	&	2FGL J1309.3+1154	&	 4C +12.46	&	 	bzb	&	LSP	&	N-06	&	\nodata	&	\nodata	\\
J13104+3220\tablenotemark{M}	&	 	1FGL J1310.6+3222	&	2FGL J1310.6+3222	&	 B2 1308+32	&	 	bzq	&	LSP	&	M-11	&	10684	&	4824	\\
J13127+4828	&	 	1FGL J1312.4+4827	&	2FGL J1312.8+4828	&	 CGRaBS J1312+4828	&	 	bzq	&	LSP	&	2LAC	&	31078	&	75540	\\
J13147+2348	&	 	1FGL J1314.7+2346	&	2FGL J1314.6+2348	&	 CRATES J1314+2348	&	 	bzb	&	HSP	&	N-06	&	\nodata	&	\nodata	\\
J13176+3425	&	 	1FGL J1317.8+3425	&	2FGL J1317.9+3426	&	 B2 1315+34A	&	 	bzq	&	LSP	&	2LAC	&	8214	&	3038	\\
J13211+2216	&	 	1FGL J1321.1+2214	&	\nodata	&	 CGRaBS J1321+2216	&	 	bzq	&	LSP	&	fit	&	11720	&	\nodata	\\
F13213+8310	&	 	1FGL J1321.3+8310	&	2FGL J1322.6+8313	&	 CRATES J1321+8316	&	 	agu	&	\nodata	&	\nodata	&	13852	&	8286	\\
J13270+2210	&	 	1FGL J1326.6+2213	&	2FGL J1326.8+2210	&	 B2 1324+22	&	 	bzq	&	LSP	&	M-11	&	6870	&	6316	\\
J13307+5202	&	 	1FGL J1331.0+5202	&	\nodata	&	 CGRaBS J1330+5202	&	 	agn	&	LSP	&	fit	&	24849	&	\nodata	\\
J13327+4722	&	 	1FGL J1332.9+4728	&	2FGL J1332.7+4725	&	 B3 1330+476	&	 	bzq	&	LSP	&	2LAC	&	9111	&	5082	\\
J13338+5057	&	 	1FGL J1333.2+5056	&	2FGL J1333.5+5058	&	 CLASS J1333+5057	&	 	agu	&	LSP	&	2LAC	&	186614	&	79197	\\
J13455+4452	&	 	1FGL J1345.4+4453	&	2FGL J1345.4+4453	&	 B3 1343+451	&	 	bzq	&	LSP	&	2LAC	&	41634	&	34870	\\
J13508+3034	&	 	1FGL J1351.0+3035	&	2FGL J1350.8+3035\tablenotemark{**}	&	 B2 1348+30B	&	 	bzq	&	LSP	&	2LAC	&	4955	&	5533	\\
F13533+1434	&	 	1FGL J1353.3+1434	&	2FGL J1353.3+1435	&	 PKS 1350+148	&	 	bzb	&	LSP	&	2LAC	&	\nodata	&	\nodata	\\
F13581+7646	&	 	1FGL J1358.1+7646	&	2FGL J1358.1+7644	&	 CGRaBS J1357+7643	&	 	bzq	&	LSP	&	2LAC	&	11881	&	5137	\\
J13590+5544	&	 	1FGL J1359.1+5539	&	2FGL J1359.4+5541	&	 CRATES J1359+5544	&	 	bzq	&	LSP	&	fit	&	42718	&	22483	\\
J14270+2348\tablenotemark{M}	&	 	1FGL J1426.9+2347	&	2FGL J1427.0+2347	&	 PKS 1424+240	&	 	bzb	&	HSP	&	N-06	&	\nodata	&	\nodata	\\
J14340+4203	&	 	1FGL J1433.9+4204	&	2FGL J1433.8+4205	&	 B3 1432+422	&	 	bzq	&	LSP	&	fit	&	20675	&	12034	\\
J14366+2321	&	 	1FGL J1436.9+2314	&	2FGL J1436.9+2319	&	 PKS 1434+235	&	 	bzq	&	LSP	&	2LAC	&	4730	&	6099	\\
J14388+3710	&	 	1FGL J1438.7+3711\tablenotemark{*}	&	2FGL J1438.7+3712\tablenotemark{**}	&	 B2 1436+37B	&	 	bzq	&	LSP	&	2LAC	&	24552	&	20208	\\
F14387+3711	&	 	1FGL J1438.7+3711\tablenotemark{*}	&	2FGL J1438.7+3712\tablenotemark{**}	&	 CRATES J1439+3712	&	 	bzq	&	LSP	&	2LAC	&	96541	&	81893	\\
F14438+2457	&	 	1FGL J1443.8+2457	&	2FGL J1444.1+2500	&	 PKS 1441+25	&	 	bzq	&	LSP	&	2LAC	&	6285	&	5206	\\
J14509+5201	&	 	1FGL J1451.0+5204	&	\nodata	&	 CLASS J1450+5201	&	 	bzb	&	\nodata	&	\nodata	&	\nodata	&	\nodata	\\
J14544+5124	&	 	1FGL J1454.6+5125	&	2FGL J1454.4+5123	&	 CRATES J1454+5124	&	 	bzb	&	ISP	&	2LAC	&	59853	&	32796	\\
F15044+1029\tablenotemark{M}	&	 	1FGL J1504.4+1029	&	2FGL J1504.3+1029	&	 PKS 1502+106	&	 	bzq	&	LSP	&	M-11	&	150686	&	64921	\\
J15061+3730	&	 	1FGL J1505.8+3725	&	2FGL J1506.0+3729	&	 B2 1504+37	&	 	bzq	&	LSP	&	fit	&	4451	&	2347	\\
J15169+1932\tablenotemark{M}	&	 	1FGL J1516.9+1928	&	2FGL J1516.9+1925	&	 PKS 1514+197	&	 	bzb	&	LSP	&	M-11	&	5431	&	3624	\\
F15197+4216	&	 	1FGL J1519.7+4216	&	2FGL J1520.9+4209	&	 B3 1518+423	&	 	bzq	&	\nodata	&	\nodata	&	37320	&	20583	\\
J15221+3144\tablenotemark{M}	&	 	1FGL J1522.1+3143	&	2FGL J1522.1+3144	&	 B2 1520+31	&	 	bzq	&	LSP	&	2LAC	&	132237	&	95093	\\
J15396+2744	&	 	1FGL J1539.7+2747	&	2FGL J1539.5+2747	&	 CGRaBS J1539+2744	&	 	bzq	&	LSP	&	2LAC	&	12644	&	8504	\\
J15429+6129\tablenotemark{M}	&	 	1FGL J1542.9+6129	&	2FGL J1542.9+6129	&	 CRATES J1542+6129	&	 	bzb	&	ISP	&	M-11	&	\nodata	&	\nodata	\\
F15534+1255\tablenotemark{M}	&	 	1FGL J1553.4+1255	&	2FGL J1553.5+1255	&	 PKS 1551+130	&	 	bzq	&	LSP	&	fit	&	21961	&	7577	\\
F15557+1111\tablenotemark{M}	&	 	1FGL J1555.7+1111	&	2FGL J1555.7+1111	&	 PG 1553+113	&	 	bzb	&	HSP	&	N-06	&	22277	&	17485	\\
J16046+5714	&	 	1FGL J1604.3+5710	&	2FGL J1604.6+5710	&	 CGRaBS J1604+5714	&	 	bzq	&	LSP	&	2LAC	&	13645	&	10693	\\
J16071+1551	&	 	1FGL J1607.1+1552	&	2FGL J1607.0+1552	&	 4C +15.54	&	 	agn	&	LSP	&	M-11	&	6887	&	5501	\\
F16090+1031	&	 	1FGL J1609.0+1031	&	2FGL J1608.5+1029	&	 4C +10.45	&	 	bzq	&	LSP	&	M-11	&	8528	&	5523	\\
J16136+3412\tablenotemark{M}	&	 	1FGL J1613.5+3411	&	2FGL J1613.4+3409	&	 B2 1611+34	&	 	bzq	&	LSP	&	M-11	&	907	&	591	\\
J16160+4632	&	 	1FGL J1616.1+4637	&	\nodata	&	 CRATES J1616+4632	&	 	bzq	&	LSP	&	fit	&	61224	&	\nodata	\\
F16302+5220	&	 	1FGL J1630.2+5220	&	2FGL J1630.4+5218	&	 CRATES J1630+5221	&	 	bzb	&	ISP	&	2LAC	&	\nodata	&	\nodata	\\
F16354+8228	&	 	1FGL J1635.4+8228	&	2FGL J1629.4+8236	&	 NGC 6251	&	 	agn	&	LSP	&	M-11	&	2942	&	983	\\
J16377+4717	&	 	1FGL J1637.9+4707	&	2FGL J1637.7+4714	&	 4C +47.44	&	 	bzq	&	LSP	&	M-11	&	5891	&	5219	\\
F16410+1143	&	 	1FGL J1641.0+1143	&	2FGL J1641.0+1141	&	 CRATES J1640+1144	&	 	agn	&	\nodata	&	\nodata	&	17669	&	7987	\\
J16475+4950	&	 	1FGL J1647.4+4948	&	2FGL J1647.5+4950	&	 CGRaBS J1647+4950	&	 	agn	&	ISP	&	2LAC	&	8867	&	8153	\\
J16568+6012	&	 	1FGL J1656.9+6017	&	2FGL J1656.5+6012	&	 CRATES J1656+6012	&	 	bzq	&	LSP	&	2LAC	&	3957	&	3412	\\
F17001+6830\tablenotemark{M}	&	 	1FGL J1700.1+6830	&	2FGL J1700.2+6831	&	 CGRaBS J1700+6830	&	 	bzq	&	LSP	&	2LAC	&	8789	&	11526	\\
J17096+4318	&	 	1FGL J1709.6+4320	&	2FGL J1709.7+4319	&	 B3 1708+433	&	 	bzq	&	LSP	&	2LAC	&	31268	&	27091	\\
F17192+1745\tablenotemark{M}	&	 	1FGL J1719.2+1745	&	2FGL J1719.3+1744	&	 PKS 1717+177	&	 	bzb	&	LSP	&	M-11	&	3062	&	1129	\\
F17225+1012	&	 	1FGL J1722.5+1012	&	2FGL J1722.7+1013	&	 CRATES J1722+1013	&	 	bzq	&	LSP	&	2LAC	&	12070	&	7465	\\
J17240+4004	&	 	1FGL J1724.0+4002	&	2FGL J1724.0+4003	&	 B2 1722+40	&	 	agn	&	LSP	&	N-06	&	12010	&	9989	\\
F17250+1151\tablenotemark{M}	&	 	1FGL J1725.0+1151	&	2FGL J1725.0+1151	&	 CGRaBS J1725+1152	&	 	bzb	&	HSP	&	N-06	&	32112	&	27437	\\
J17274+4530\tablenotemark{M}	&	 	1FGL J1727.3+4525	&	2FGL J1727.1+4531	&	 B3 1726+455	&	 	bzq	&	LSP	&	M-11	&	3774	&	4284	\\
J17283+5013	&	 	1FGL J1727.9+5010	&	2FGL J1728.2+5015	&	 I Zw187	&	 	bzb	&	HSP	&	M-11	&	8874	&	4417	\\
F17308+3716	&	 	1FGL J1730.8+3716	&	2FGL J1731.3+3718	&	 CRATES J1730+3714	&	 	bzb	&	ISP	&	2LAC	&	\nodata	&	\nodata	\\
J17343+3857\tablenotemark{M}	&	 	1FGL J1734.4+3859	&	2FGL J1734.3+3858	&	 B2 1732+38A	&	 	bzq	&	LSP	&	2LAC	&	10666	&	5878	\\
J17425+5945	&	 	1FGL J1742.1+5947	&	2FGL J1742.1+5948	&	 CRATES J1742+5945	&	 	bzb	&	ISP	&	M-11	&	\nodata	&	\nodata	\\
F17442+1934	&	 	1FGL J1744.2+1934	&	2FGL J1744.1+1934	&	 1ES 1741+196	&	 	bzb	&	HSP	&	N-06	&	4098	&	2228	\\
F17485+7004	&	 	1FGL J1748.5+7004	&	2FGL J1748.8+7006	&	 CGRaBS J1748+7005	&	 	bzb	&	LSP	&	M-11	&	3660	&	2589	\\
J17490+4321	&	 	1FGL J1749.0+4323	&	2FGL J1749.1+4323	&	 B3 1747+433	&	 	bzb	&	LSP	&	M-11	&	\nodata	&	\nodata	\\
F17566+5524	&	 	1FGL J1756.6+5524\tablenotemark{*}	&	\nodata	&	 CRATES J1757+5523	&	 	agn	&	LSP	&	fit	&	28753	&	13989	\\
F18004+7827\tablenotemark{M}	&	 	1FGL J1800.4+7827	&	2FGL J1800.5+7829	&	 CGRaBS J1800+7828	&	 	bzb	&	LSP	&	M-11	&	2119	&	2160	\\
F18070+6945\tablenotemark{M}	&	 	1FGL J1807.0+6945	&	2FGL J1806.7+6948	&	 3C 371	&	 	bzb	&	ISP	&	M-11	&	2720	&	1674	\\
F18096+2908	&	 	1FGL J1809.6+2908	&	2FGL J1809.7+2909	&	 CRATES J1809+2910	&	 	bzb	&	\nodata	&	\nodata	&	\nodata	&	\nodata	\\
F18134+3141	&	 	1FGL J1813.4+3141	&	2FGL J1813.5+3143	&	 B2 1811+31	&	 	bzb	&	HSP	&	N-06	&	15967	&	10872	\\
F18240+5651\tablenotemark{M}	&	 	1FGL J1824.0+5651	&	2FGL J1824.0+5650	&	 4C +56.27	&	 	bzb	&	LSP	&	M-11	&	5013	&	4875	\\
F18485+3224\tablenotemark{M}	&	 	1FGL J1848.5+3224	&	2FGL J1848.5+3216	&	 B2 1846+32A	&	 	bzq	&	LSP	&	2LAC	&	21979	&	21036	\\
F18493+6705\tablenotemark{M}	&	 	1FGL J1849.3+6705	&	2FGL J1849.4+6706	&	 CGRaBS J1849+6705	&	 	bzq	&	LSP	&	A-11	&	13397	&	5139	\\
F18525+4853	&	 	1FGL J1852.5+4853	&	2FGL J1852.5+4856	&	 CGRaBS J1852+4855	&	 	bzq	&	LSP	&	2LAC	&	16991	&	11696	\\
F19030+5539\tablenotemark{M}	&	 	1FGL J1903.0+5539	&	2FGL J1903.3+5539	&	 CRATES J1903+5540	&	 	bzb	&	ISP	&	N-06	&	\nodata	&	\nodata	\\
F19416+7214	&	 	1FGL J1941.6+7214	&	2FGL J1941.6+7218	&	 CRATES J1941+7221	&	 	agu	&	\nodata	&	\nodata	&	\nodata	&	\nodata	\\
F20000+6508\tablenotemark{M}	&	 	1FGL J2000.0+6508	&	2FGL J2000.0+6509	&	 1ES 1959+650	&	 	bzb	&	HSP	&	N-06	&	14530	&	10334	\\
F20019+7040	&	 	1FGL J2001.9+7040	&	2FGL J2001.7+7042	&	 CRATES J2001+7040	&	 	agu	&	LSP	&	2LAC	&	\nodata	&	\nodata	\\
F20060+7751	&	 	1FGL J2006.0+7751	&	2FGL J2004.5+7754	&	 CGRaBS J2005+7752	&	 	bzb	&	LSP	&	M-11	&	3248	&	1172	\\
F20091+7228	&	 	1FGL J2009.1+7228	&	2FGL J2009.7+7225	&	 4C +72.28	&	 	bzb	&	LSP	&	N-06	&	\nodata	&	\nodata	\\
F20204+7608	&	 	1FGL J2020.4+7608	&	2FGL J2022.5+7614	&	 CGRaBS J2022+7611	&	 	bzb	&	ISP	&	N-06	&	\nodata	&	\nodata	\\
F20315+1219\tablenotemark{M}	&	 	1FGL J2031.5+1219	&	2FGL J2031.7+1223	&	 PKS 2029+121	&	 	bzb	&	LSP	&	2LAC	&	5603	&	2629	\\
F20354+1100	&	 	1FGL J2035.4+1100	&	2FGL J2035.4+1058	&	 PKS 2032+107	&	 	bzq	&	LSP	&	N-06	&	12343	&	7609	\\
F20497+1003	&	 	1FGL J2049.7+1003\tablenotemark{*}	&	2FGL J2049.8+1001	&	 PKS 2047+098	&	 	agu	&	\nodata	&	\nodata	&	\nodata	&	\nodata	\\
F21155+2937	&	 	1FGL J2115.5+2937	&	2FGL J2115.3+2932	&	 B2 2113+29	&	 	bzq	&	LSP	&	2LAC	&	8184	&	4537	\\
F21161+3338	&	 	1FGL J2116.1+3338	&	2FGL J2116.2+3339	&	 B2 2114+33	&	 	bzb	&	ISP	&	2LAC	&	\nodata	&	\nodata	\\
F21209+1901	&	 	1FGL J2120.9+1901	&	2FGL J2121.0+1901	&	 OX 131	&	 	bzq	&	LSP	&	2LAC	&	20415	&	14131	\\
F21434+1742\tablenotemark{M}	&	 	1FGL J2143.4+1742	&	2FGL J2143.5+1743	&	 OX 169	&	 	bzq	&	LSP	&	M-11	&	13782	&	10365	\\
F21525+1734	&	 	1FGL J2152.5+1734	&	2FGL J2152.4+1735	&	 PKS 2149+17	&	 	bzb	&	LSP	&	N-06	&	3435	&	3951	\\
F21574+3129	&	 	1FGL J2157.4+3129	&	2FGL J2157.4+3129	&	 B2 2155+31	&	 	bzq	&	LSP	&	2LAC	&	17937	&	16558	\\
F22035+1726\tablenotemark{M}	&	 	1FGL J2203.5+1726	&	2FGL J2203.4+1726	&	 PKS 2201+171	&	 	bzq	&	LSP	&	M-11	&	15029	&	10498	\\
F22121+2358	&	 	1FGL J2212.1+2358	&	2FGL J2211.9+2355	&	 PKS 2209+236	&	 	bzq	&	LSP	&	2LAC	&	2230	&	1340	\\
F22171+2423	&	 	1FGL J2217.1+2423	&	2FGL J2217.1+2422	&	 B2 2214+24B	&	 	bzb	&	LSP	&	2LAC	&	5967	&	3161	\\
F22193+1804	&	 	1FGL J2219.3+1804	&	2FGL J2219.1+1805	&	 CGRaBS J2219+1806	&	 	bzq	&	\nodata	&	\nodata	&	20104	&	7251	\\
F22362+2828\tablenotemark{M}	&	 	1FGL J2236.2+2828	&	2FGL J2236.4+2828	&	 B2 2234+28A	&	 	bzq	&	LSP	&	M-11	&	5173	&	2914	\\
F22440+2021\tablenotemark{M}	&	 	1FGL J2244.0+2021	&	2FGL J2243.9+2021	&	 CRATES J2243+2021	&	 	bzb	&	HSP	&	2LAC	&	\nodata	&	\nodata	\\
F22501+3825	&	 	1FGL J2250.1+3825	&	2FGL J2250.0+3825	&	 B3 2247+381	&	 	bzb	&	HSP	&	N-06	&	18920	&	13435	\\
F22517+4030	&	 	1FGL J2251.7+4030	&	2FGL J2251.9+4032	&	 CRATES J2251+4030	&	 	bzb	&	\nodata	&	\nodata	&	\nodata	&	\nodata	\\
F22539+1608\tablenotemark{M}	&	 	1FGL J2253.9+1608	&	2FGL J2253.9+1609	&	 3C 454.3	&	 	bzq	&	LSP	&	M-11	&	12137	&	15731	\\
F23073+1452	&	 	1FGL J2307.3+1452	&	2FGL J2308.0+1457	&	 CGRaBS J2307+1450	&	 	bzb	&	LSP	&	fit	&	30212	&	40301	\\
F23110+3425	&	 	1FGL J2311.0+3425	&	2FGL J2311.0+3425	&	 B2 2308+34	&	 	bzq	&	LSP	&	2LAC	&	12526	&	12080	\\
F23220+3208	&	 	1FGL J2322.0+3208	&	2FGL J2322.2+3206	&	 B2 2319+31	&	 	bzq	&	LSP	&	2LAC	&	10989	&	4821	\\
F23216+2726	&	 	1FGL J2321.6+2726	&	2FGL J2321.0+2737	&	 4C +27.50	&	 	bzq	&	LSP	&	2LAC	&	5914	&	2546	\\
F23226+3435	&	 	1FGL J2322.6+3435	&	2FGL J2322.6+3435	&	 CRATES J2322+3436	&	 	bzb	&	HSP	&	N-06	&	52497	&	13883	\\
F23252+3957	&	 	1FGL J2325.2+3957	&	2FGL J2325.3+3957	&	 B3 2322+396	&	 	bzb	&	HSP	&	N-06	&	\nodata	&	\nodata	\\
\enddata
\tablecomments{Col.\ (1): Source Name: if name starts with 'J' it is a VIPS or pre-1FGL MOJAVE source; if name starts with 'F' it is a source targeted specifically for its presence in 1FGL.
Col.\ (2): 1FGL source name.  
Col.\ (3): 2FGL source name.
Col.\ (4): Other AGN name.
Col.\ (5): Optical Type from 2LAC/1LAC: bzb = BL Lac object, bzq = FSRQ, agn = non-blazar AGN, agu = AGN of uncertain type, sbg = starburst galaxy.
Col.\ (6): SED type: LSP=low-synchrotron peaked, ISP=intermediate-synchrotron peaked, HSP=high-synchrotron peaked. 
Col.\ (7): Reference for SED type: 0FGL = Abdo et al. 2010c, N-06 = Nieppola et a. 2006, N-08 = Nieppola et al. 2008, A-11 = Aatrokoski et al. 2011, M-11 = Meyer et al. 2011, 2LAC = estimated using technique in Ackermann et al. 2011, fit = log-parabolic fit to NED data.
Col.\ (8): Gamma-ray loudness using 1FGL data.
Col.\ (9): Gamma-ray loudness using 2FGL data.
}
\tablenotetext{M}{MOJAVE source that was included in Lister et al. (2011). For more data, visit the MOJAVE website http://www.physics.purdue.edu/astro/MOJAVE/MOJAVEIItable.html or see Lister et al. (2009b, 2011)}
\tablenotetext{*}{Indicates a LAT source which is associated with multiple radio sources with high ($\geq$80\%) probability in 1LAC}
\tablenotetext{**}{Indicates a LAT source which is associated with multiple radio sources with high ($\geq$80\%) probability in 2LAC}
\tablenotetext{c}{Indicates a 2FGL source that is considered to be potentially confused with galactic diffuse emission}
\label{datatable}
\end{deluxetable}

\textwidth = 7.0truein
\textheight = 10.0truein
\begin{deluxetable}{ccrr}
\tablecolumns{4}
\tabletypesize{\scriptsize}
\tablewidth{0pt}
\tablecaption{BL Lac Object $G_{r} - \nu^{S}_{peak}$ Correlations}
\tablehead{
\colhead{LAT data} & \colhead{BL Lac Sample} & \colhead{Spearman $\rho_{S}$}	&	\colhead{Spearman $p$}\\}
\startdata
1FGL & Original BL Lac Object Sample (27 LBLs)	&	0.46	&	0.0003\\
 & Adding 7 LSP FSRQs	&	0.42	&	0.0005\\
 & Adding 14 LSP FSRQs	&	0.37	&	0.001\\
 & Adding 27 LSP FSRQs	&	0.30	&	0.005\\
 & Adding 54 LSP FSRQs	&	0.17	&	0.07\\
 & Adding all 96 LSP FSRQs	&	0.14	&	0.08\\
 & & & \\
2FGL & Original BL Lac Object Sample (26 LBLs) &	0.40	&	0.003\\
 & Adding 7 LSP FSRQs	&	0.30	&	0.02\\
 & Adding 13 LSP FSRQs	&	0.30	&	0.01\\
 & Adding 26 LSP FSRQs	&	0.25	&	0.02\\
 & Adding 52 LSP FSRQs	&	0.13	&	0.2\\
 & Adding all 91 LSP FSRQs	&	0.12	&	0.2\\
\enddata
\label{phantomtable}
\end{deluxetable}

\clearpage

\clearpage

\clearpage

\clearpage

\clearpage

\clearpage

\clearpage

\clearpage

\clearpage

\clearpage

\clearpage

\clearpage

\end{document}